\documentclass[11pt]{article}
\usepackage[margin=1in]{geometry}
\usepackage[inline,shortlabels]{enumitem}
\usepackage{mathtools}
\usepackage{subdepth}
\usepackage{graphicx}
\usepackage[english]{babel}
\usepackage{color}
\usepackage{amssymb,amsmath,amsthm}
\usepackage{bbm}
\usepackage[OT1]{fontenc}
\usepackage{booktabs}
\usepackage{authblk}
\usepackage[round]{natbib}

\makeatletter
\renewcommand\section{\@startsection{section}{1}{\z@}%
  {-3.5ex \@plus -1ex \@minus -.2ex}%
  {2.3ex \@plus.2ex}%
  {\normalfont\Large\bfseries\raggedright}}
\renewcommand\subsection{\@startsection{subsection}{2}{\z@}%
  {-3.25ex\@plus -1ex \@minus -.2ex}%
  {1.5ex \@plus .2ex}%
  {\normalfont\large\bfseries\raggedright}}
\renewcommand\subsubsection{\@startsection{subsubsection}{3}{\z@}%
  {-3.25ex\@plus -1ex \@minus -.2ex}%
  {1.5ex \@plus .2ex}%
  {\normalfont\normalsize\bfseries\raggedright}}
\makeatother

\usepackage[colorlinks,citecolor=blue,linkcolor=blue,urlcolor=blue]{hyperref}
\usepackage{orcidlink}

\DeclareMathOperator{\var}{\mathsf{Var}}

\DeclarePairedDelimiterX{\norm}[1]{\lVert}{\rVert}{#1}

\def\keywordfont{\fontsize{9}{10}\selectfont}
\newenvironment{keywords}{\par\addvspace{8pt}%
                          \keywordfont\noindent{\itshape Keywords}:\ \ignorespaces%
                         }{\par\addvspace{23pt}}

\newtheorem{theorem}{Theorem}
\newtheorem{lemma}{Lemma}
\newtheorem{prop}{Proposition}

\newtheorem{remark}{Remark}

\theoremstyle{definition}
\newtheorem{assumption}{Assumption}
\newtheorem{definition}{Definition}
\theoremstyle{plain}

\newcommand{\titlepaper}{Efficient transport and generalization of survival treatment effects}

\date{}
\title{\titlepaper}
\author[1]{Axel Martin\,\orcidlink{0000-0002-4665-9113}%
\thanks{Corresponding author: asm8744@nyu.edu}}
\author[1]{Iv\'an D\'iaz\,\orcidlink{0000-0001-9056-2047}}
\author[1]{Michele Santacatterina\,\orcidlink{0000-0003-0599-3876}}

\affil[1]{\small Division of Biostatistics, Department of Population Health,
New York University Grossman School of Medicine,
180 Madison Avenue, New York, NY 10016, U.S.A.}

\renewcommand{\P}{P}
\newcommand{\Po}{P_0}
\newcommand{\Pn}{P_n}

\newcommand{\PP}{\mathbb{P}}
\newcommand{\PPn}{\mathbb{P}_n}
\newcommand{\PPo}{\mathbb{P}_o}
\newcommand{\G}{G}
\newcommand{\Go}{G_0}

\renewcommand{\S}{S}
\newcommand{\So}{S_0}

\newcommand{\E}{\mathsf{E}}
\newcommand{\Eo}{\mathsf{E}_0}
\newcommand{\Ee}{\mathsf{E}_{\epsilon}}

\newcommand{\EE}{\mathbb{E}}

\newcommand{\f}{\mathsf{f}}
\newcommand{\fo}{\mathsf{f}_0}

\newcommand{\g}{\mathsf{g}}

\newcommand{\Mset}{\mathcal{M}}

\renewcommand{\L}{\mathsf{L}}

\newcommand{\1}{\mathbbm{1}}

\newcommand{\indep}{\mbox{$\perp\!\!\!\perp$}}
\newcommand{\dd}{\mathsf{d}}
\newcommand{\ppe}{\frac{\partial}{ \partial \epsilon}}
\newcommand{\ate}{\mid_{\epsilon = 0}}
\newcommand{\Se}{S_{\epsilon}}
\newcommand{\Ge}{G_{\epsilon}}
\newcommand{\Pe}{P_{\epsilon}}
\newcommand{\he}{h_{\epsilon}}
\newcommand{\hprod}{\prod_{m = 1}^t}
\newcommand{\hsum}{\sum_{m = 1}^t}
\newcommand{\hsumtau}{\sum_{m = 1}^{\tau}}
\newcommand{\D}{\mathsf{D}}
\newcommand{\R}{\mathsf{R}}
\newcommand{\p}{\mathsf{p}}

\begin{document}
\label{firstpage}

\maketitle

\begin{abstract}
Randomized controlled trials provide internally valid
estimates of treatment effects, but their results
may not directly apply to broader target populations
due to differences in baseline covariate distributions,
adherence to treatment or variations in outcome mechanisms.
Under standard transport and time-to-event identifiability
assumptions, we develop nonparametric,
debiased machine learning estimators
for transporting and generalizing causal survival
treatment effect differences from a source population
to a target population in discrete time.
We derive the efficient influence functions for
the transport and generalization survival difference
estimands and propose cross-fitted one-step estimators
that are doubly robust and achieve semiparametric
efficiency bounds under weak regularity conditions.
We further introduce estimators that exploit
known effect modifier subsets through an additive
parameterization of the survival function,
reducing the dimensionality of the reweighting
and yielding smaller or equal asymptotic variance.
We establish asymptotic normality, double robustness,
and rates of convergence for all proposed estimators.
Finite-sample properties are illustrated through
Monte Carlo simulations under flexible and
misspecified nuisance estimation scenarios.
We apply the methods to data from the
Women's Health Initiative to estimate the
effect of hormone therapy on coronary heart disease
across trial and observational populations.
\end{abstract}


\section{Introduction}

Randomized controlled trials (RCTs) remain the
gold standard for estimating causal treatment effects,
owing to the elimination of confounding
through random treatment assignment.
However, trial populations often differ systematically
from the broader populations to which
their findings are intended to apply
\citep{stuart2011use, cole2010generalizing}.
Strict eligibility criteria, volunteer bias, and
logistical constraints of trial enrollment
produce study samples that may not reflect
the covariate distributions of real-world
target populations \citep{degtiar2023review}.
When treatment effect heterogeneity is present,
these distributional differences can render
naive extrapolation of trial results misleading,
motivating formal methods for
transporting and generalizing causal effects
across populations.

A growing body of work has addressed the
problem of extending causal inferences from
a source population to a target population.
\citet{pearl2011transportability} and
\citet{bareinboim2016causal} formalized
transportability within the structural causal
model framework, establishing graphical
conditions under which causal effects identified
in one population can be validly transferred
to another.
In parallel, a statistical literature has
developed weighting and outcome modeling
approaches for this problem.
\citet{cole2010generalizing} and
\citet{buchanan2018generalizing} proposed
inverse probability of sampling weights
to generalize trial findings,
while \citet{stuart2011use} and
\citet{tipton2013improving} introduced
propensity score methods for assessing
and improving generalizability.
\citet{westreich2017transportability} developed
inverse odds of sampling weights
for transporting treatment effects.
\citet{dahabreh2019generalizing} established
a potential outcomes framework for
generalizing from randomized trials
to all trial-eligible individuals, deriving
identification results and proposing
doubly robust estimators for binary
and continuous outcomes.
Subsequent work extended these ideas to
multiple trials \citep{dahabreh2023efficient},
non-nested designs \citep{li2022generalizing},
and various study design considerations
\citep{dahabreh2021study, lesko2017generalizing}.
\citet{rudolph2017robust} developed targeted
learning estimators for transporting
encouragement design effects across sites.
\citet{degtiar2023review} provides a
comprehensive review of this literature.

Despite this progress, the existing
generalizability and transportability methods
have been developed primarily for
binary or continuous outcomes.
Time-to-event outcomes, which arise naturally
in clinical trials evaluating treatments
for chronic diseases, cancer, and cardiovascular
conditions, introduce additional complexities
due to right censoring, time-varying
risk sets, and the need to model
discrete or continuous hazard functions.
Causal survival analysis has a rich
methodological tradition rooted in the
work of \citet{Robins86, Robins87}
on g-computation and structural models
for censored outcomes.
Semiparametric efficient estimators for
survival parameters have been developed
using targeted minimum loss-based estimation
\citep{vanderLaanRose11, Stitelman2011}
and one-step debiased approaches
\citep{diaz2019statistical,
kennedy2023semiparametricdoublyrobusttargeted}.
Cross-fitting techniques
\citep{chernozhukov2018double, zheng2011cross}
enable the use of flexible machine
learning algorithms for nuisance estimation
while preserving valid inference.
However, the intersection of transportability
with survival outcomes has received
limited attention.
\citet{lee2022doubly} developed doubly robust, locally efficient
estimators for the survival treatment effect in
a target population over a broad class
of functionals of the treatment-specific survival curves,
using augmented calibration weighting with Cox working
models and a method-of-sieves extension. For the
survival-difference functional their target parameter coincides with
our transport estimand, so in that case
the two share the same estimand. The
estimators differ in construction: they operate in
continuous time and attain flexibility through penalized
power-series sieves under Donsker-type regularity conditions, without
sample splitting; we instead work in discrete
time and estimate the nuisances with cross-fitting,
which removes the Donsker restriction, accommodates arbitrary
data-adaptive learners, and attains the nonparametric efficiency
bound under rate conditions. \citet{cao2024transporting}
likewise study transport of survival effects to
a target population, developing inverse probability weighted
and doubly robust estimators of the counterfactual
survival functions together with approximate variance estimators,
again in continuous time and under parametric
working models. Neither of these works exploits
structural knowledge about effect modification for efficiency
gains or treats the generalization estimand; we
additionally define a generalization estimand and introduce
structured estimators that leverage a known effect-modifier
subset to reduce the dimensionality of the
reweighting and lower the asymptotic variance.

In many applied settings, substantive
knowledge identifies a subset of
covariates that modify the treatment effect.
When these effect modifiers are known,
one can impose structure on the
outcome model to reduce the
dimensionality of the reweighting problem.
This idea connects to a broader
theme in semiparametric theory:
restricting the statistical model
shrinks the nuisance tangent space,
which can only reduce the
semiparametric efficiency bound
\citep{Bickel97, tsiatis2006semiparametric}.
We adapt this construction
to the transportability and generalizability
with time-to-event outcomes setting.
Exploiting such structure is particularly
valuable in transportability problems,
where the reweighting over the full
covariate space $W$ can suffer from
high variance due to near-violations
of the positivity assumption.
If the conditional treatment effect depends
on $W$ only through a
lower-dimensional subset $V$,
and population differences in
$V$ are driven by an
even smaller subset $Z \subseteq V$,
then reweighting over $Z$ alone
suffices and yields
substantially more stable estimators.

In this paper, we develop nonparametric
debiased machine learning estimators
for causal survival treatment
effect differences in the
context of generalizability and transportability.
We focus on the common setting in which
outcome data are available only in the
source (trial) population, while the
target population contributes only
baseline covariate information.
After defining transport and generalization
survival difference estimands in
discrete time and derive their
efficient influence functions
under a nonparametric model, we make two
major contributions.
First, we propose structured versions
of these estimands that exploit known
effect modifier subsets through
an additive parameterization of the
survival function,
combined with a partial
heterogeneity condition linking
population membership to
the effect modifiers.
We derive the efficient influence
functions for these structured estimands
and show that their semiparametric
variance bounds are no larger
than those of their unstructured
counterparts.
Second, we establish the
double robustness, asymptotic normality,
and rates of convergence for
all estimators, implemented
via cross-fitted one-step estimation
with flexible data-adaptive
nuisance estimators.
We evaluate finite-sample
properties through Monte Carlo
simulations and illustrate the methods
using data from the Women's Health
Initiative (WHI)
\citep{rossouw2002risks, prentice2005combined},
where we estimate the effect of
hormone therapy on coronary heart
disease across trial and observational
populations.


\section{Notation, causal assumptions and estimands} \label{sec:id}

Consider $n$ individuals monitored at
$K$ discrete time points
$t \in \{1, \dots, K\}$.
Let $T$ denote a time-to-event outcome
taking values in
$\{1, \dots, K\} \cup \{\infty\}$,
where $T = \infty$ indicates no event
within the follow-up period.
Let $C \in \{0, \dots, K\}$ denote the
censoring time, with $C = K$ representing
administrative censoring.
Let $A \in \{0,1\}$ denote binary treatment
assignment, $\Gamma \in \{0,1\}$ the population
membership indicator with $\Gamma = 1$
for the source population and $\Gamma = 0$
for the target population, and
$W \in \mathbb{R}^d$ the vector of
baseline covariates.
We define the potential outcomes
$T_a$, $a \in \{0,1\}$, as the event times
that would have been observed under
treatment $A = a$ and censoring
$C = K$ with probability one.
For a restriction time
$\tau \in \{1, \dots, K\}$,
define the restricted survival time
under arm $a$ as $\min\{T_a, \tau\}$.

We have $\check{T} = \min(C, T)$
as the observed follow-up time and
$\Delta = \1(T \leq C)$ as the event
indicator.
The observable data consist of
$O = (W, A, \Gamma, \Gamma\Delta, \Gamma\check{T})$,
where $\Gamma\Delta$ and $\Gamma\check{T}$ denote that these
variables are only observed in the source population.
We assume that observations
$O_1, \dots, O_n$ are independent
and identically distributed
according to an unknown distribution
$\Po$, with empirical distribution $\Pn$.
Let $\Po \in \Mset$, where $\Mset$
is the nonparametric model defined
as all continuous densities on $O$
with respect to a dominating measure $v$.
Let $\P$ denote a generic distribution
in $\Mset$.
We write $\Eo[\cdot]$ for the
expectation with respect to $\Po$
and $\EE$ for the expectation
over draws of $O_1, \dots, O_n$.
For a function $\f(O)$, define
$\P \f = \int \f(O) \, \dd \P(o)$
and $\|\f\|^2 = \Po \f^2$.
We write $a \lesssim b$ to denote
that $a$ is bounded above by $b$
up to a universal constant.

This can be rewritten
as a longitudinal data structure.
Define the event and censoring indicators
$L_t = \1\{\check{T} = t, \Delta = 1\}$
and $R_t = \1\{\check{T} = t, \Delta = 0\}$
for $t \in \{0, \dots, K\}$, so that
$O = (W, A, \Gamma, \Gamma R_0, \Gamma L_1,
\Gamma R_1, \dots, \Gamma R_{\tau-1}, \Gamma L_{\tau})$.
For a random variable $X$, let
$\bar{X}_t = (X_0, \dots, X_t)$
denote its history through time $t$.
Define the at-risk indicators
$I_t = \1\{\bar{R}_{t-1} = 0, \bar{L}_{t-1} = 0\}$
and $J_t = \1\{\bar{R}_{t-1} = 0,
\bar{L}_t = 0\}$, representing
respectively whether a participant
is at risk of the event at time $t$
and at risk of censoring at time $t$,
with $J_0 = 1$.

Define the discrete survival hazard
at time $m \in \{1, \dots, K\}$ as
\[ h_0(m \mid a, w, \gamma) = \Po
\left( L_m = 1 \mid I_m = 1,
A = a, W = w, \Gamma = \gamma \right)\]
and the censoring hazard as
\[ g_{C,0}(m \mid a, w, \gamma) = \Po
\left( R_m = 1 \mid J_m = 1,
A = a, W = w, \Gamma = \gamma \right). \]
Define the propensity scores
$\pi_{A,0}(a \mid w, \gamma) =
\Po(A = a \mid W = w, \Gamma = \gamma)$
and $\pi_{\Gamma,0}(\gamma, w) =
\Po(\Gamma = \gamma \mid W = w)$.
The nuisance parameter is
$\eta = (h, \pi_A, \pi_{\Gamma}, g_C)$,
where a subscript ``$0$'' denotes
the value under $\Po$.

We now state the causal assumptions
under which the estimands of interest
are identifiable, and define
the transport and generalization
survival difference estimands.


\subsection{Transport and Generalization}

\begin{assumption}{(Conditional exchangeability)
\label{ass:condex}}
    The potential outcome is independent
    of treatment assignment given covariates
    and population membership:
    $T_a \indep A \mid W, \Gamma =0$.
\end{assumption}

\begin{assumption}{(Consistency) \label{ass:cons}}
    The observed outcome under the received
    treatment equals the corresponding
    potential outcome:
    $A = a$ implies $T_a = T$.
\end{assumption}

\begin{assumption}{(Positivity) \label{ass:pos}}
    Treatment, population membership, and
    censoring probabilities are bounded
    away from zero. For some $\epsilon > 0$, we have
    $\Po(\pi_{A,0} > \epsilon) = 1$; $\Po(\pi_{\Gamma,0} > \epsilon) = 1$;
    $\Po( \Po ( W =w \mid \Gamma = 0) > \epsilon) = 1 \implies \Po (\pi_{A,0 \mid \Gamma = 1} > \epsilon) = 1$ and $\Po (g_{C,0} (m) < 1 - \epsilon) = 1 \text{ } \forall \text{ } m \in 1, \dots , K $.
\end{assumption}
Note that only weak positivity is required for identification but that strong positivity is required for estimation.

\begin{assumption}{(Random censoring)
\label{ass:ign}}
    The survival time and
    the censoring time are conditionally
    independent given treatment, covariates,
    and population:
    $T \indep C \mid A, W, \Gamma$.
\end{assumption}

\begin{assumption}{(Transportability and
generalizability) \label{ass:trans}}
    \textbf{Transportability:} The potential survival function is
    the same across populations:
    $\Po(T_a > t \mid W = w, \Gamma = 0)
    = \Po(T_a > t \mid W = w, \Gamma = 1)$. \
    \textbf{Generalizability:} $T_a \indep \Gamma \mid W$.
\end{assumption}

The survival and censoring functions
for time $t \in \{1, \dots, \tau\}$,
treatment arm $a$, covariates $w$,
and population $\gamma$ are
\begin{equation*}
    S(t \mid a,w,\gamma) =
    P(T > t \mid A = a, W = w, \Gamma = \gamma),
    \;
    G(t \mid a,w,\gamma) =
    P(C \geq t \mid A = a, W = w, \Gamma = \gamma).
\end{equation*}
Under Assumptions~\ref{ass:condex}
through~\ref{ass:trans},
$T \indep C \mid A, W, \Gamma$, yielding
the product representations
\begin{equation*}
    S(t \mid a,w,\gamma) =
    \prod_{m = 1}^t
    \left(1 - h(m, a,w,\gamma)\right),
    \quad
    G(t \mid a,w,\gamma) =
    \prod_{m = 0}^{t-1}
    \left(1 - g_C(m, a,w,\gamma)\right).
\end{equation*}

\begin{theorem}{Estimand identification
\label{thm:id}}

    If Assumptions~\ref{ass:condex}
    through~\ref{ass:trans} hold for some
    $\tau \in (0, \infty)$, then
    $\Po(T_1 > t \mid W, \Gamma) -
    \Po(T_0 > t \mid W, \Gamma) =
    S_0(t \mid 1, W, \Gamma) -
    S_0(t \mid 0, W, \Gamma)$,
    $\Po$-almost surely for all
    $t \in [0, \tau]$, where
    $S_0(t \mid a, W, \Gamma) =
    \prod_{m=1}^t
    (1 - h_0(m, a, w, \gamma))$.

    The causal transport survival difference
    is the conditional average treatment
    effect in the target population,
    \begin{equation}\label{eq:transport}
        \theta_{T, \Po}(t) = \Eo \left[
        S_0(t \mid A = 1, W, \Gamma = 1) -
        S_0(t \mid A = 0, W, \Gamma = 1)
        \mid \Gamma = 0 \right].
    \end{equation}
    The causal generalization survival
    difference is the population-averaged
    treatment effect,
    \begin{equation}\label{eq:gen}
        \theta_{G, \Po}(t) = \Eo \left[
        S_0(t \mid A = 1, W, \Gamma = 1) -
        S_0(t \mid A = 0, W, \Gamma = 1) \right].
    \end{equation}
\end{theorem}


\subsection{Leveraging effect modifiers
for efficiency gains}

We now consider the setting in which
differences in survival between populations
arise from differences in the distributions
of baseline covariates that modify the
treatment effect.
Let $V \subseteq W$ denote a known subset
of effect modifiers, $X \subseteq W$
the covariates whose distributions differ
across populations, and $Z = V \cap X$
the effect modifiers that also differ
between populations, so that
$Z \subseteq V \subseteq W$. Specifically, assume the following.

\begin{definition}{(Additive parameterization
of the survival function) \label{ass:param}}
    The conditional survival function admits
    the decomposition
    $S(t \mid A, W, \Gamma = 1) =
    A \times \f(t,W) + \g(t,W)$.
\end{definition}

\begin{assumption}{(Treatment effect modifiers)
\label{ass:tem}}
    The function $\f(\cdot)$ depends
    on $W$ only through a known
    subvector $V = V(W)$.
\end{assumption}

\begin{assumption}{(Partial study heterogeneity
of effect modifiers) \label{ass:hem}}
    There exists a subset $Z \subseteq V$
    such that $\Gamma \perp V \mid Z$.
\end{assumption}

\begin{theorem}{Structured subset estimand
identification \label{thm:efsid}}

    If Assumptions~\ref{ass:param}
    through~\ref{ass:hem} hold in addition
    to Assumptions~\ref{ass:condex}
    through~\ref{ass:trans}, the structured
    transport survival difference is
    \begin{align}\label{eq:transport_sub}
        \lambda_{T, \Po}(t)
        &= \Eo \left[ S_0(t \mid A = 1,
        W, \Gamma = 1) - S_0(t \mid A = 0,
        W, \Gamma = 1) \mid \Gamma = 0
        \right] \notag \\
        &= \Eo \left[ \Eo \left[
        \fo(t,V) \mid Z \right]
        \mid \Gamma = 0 \right],
    \end{align}
    and the structured generalization
    survival difference is
    \begin{equation}\label{eq:gen_sub}
        \lambda_{G, \Po}(t) =
        \Eo[S_0(t \mid A = 1, W, \Gamma)
        - S_0(t \mid A = 0, W, \Gamma)]
        = \Eo[\fo(t,V)].
    \end{equation}
\end{theorem}

Identification proofs for all estimators
can be found in Web Appendix~A.


\section{Proposed Estimators}\label{sec:properties}

For the structured estimators, define the
additional propensity score
$\pi^*_{\Gamma, 0}(\gamma, z) =
\Po(\Gamma = \gamma \mid Z = z)$,
which conditions on the effect modifier
subset $Z$ rather than the full covariate
vector $W$.
The nuisance parameter for the base
estimators is
$\eta = (h, \pi_A, \pi_{\Gamma}, g_C)$,
and for the structured estimators
$\eta^* = (h, \f, \pi_A,
\pi^*_{\Gamma}, g_C)$.


\subsection{Efficient influence functions}

We derive the efficient influence functions
(EIFs) for all four estimands.
The EIF characterizes the first-order
behavior of any regular asymptotically
linear estimator and determines the
semiparametric efficiency bound
\citep{Bickel97, tsiatis2006semiparametric}.
The proofs, carried out via
Gateaux derivatives in a parametric
submodel, are given in
Web Appendix~B. In the following sections and
appendices we will omit conditional random variables 
unless required for notational convenience.

\begin{theorem}{Efficient influence
functions} \label{thm:eif}

    Under Assumptions~\ref{ass:condex}
    through~\ref{ass:trans}, if there
    exists $\epsilon > 0$ such that
    $\min\{\pi_{A,0}, \pi_{\Gamma,0},
    \Go(m)\} \geq \epsilon$ for all
    $m = 1, \dots, K$,
    $\Po$-almost everywhere with
    $\So(t \mid a, w, \gamma) > 0$. 
    Let $\p_0 = \Po(\Gamma = 0)$,
    $S^*_0(t) = \So(t \mid A = 1)
    - \So(t \mid A = 0)$, and
    \[
    \D(t) = \hsum
    \frac{\So(t)
    \, \1(I_m = 1) \,
    (h_0(m)
    - \1(L_m = 1))}
    {\So(m)
    \, \Go(m)}.
    \] 
    Then $\theta_{T,\Po}(t)$ and
    $\theta_{G,\Po}(t)$ are pathwise
    differentiable in the nonparametric
    model $\Mset$ with efficient influence
    functions. 
    \begin{align}\label{eq:trans_eif}
        \varphi_{\theta_T}(O;\eta_0, t)
        &= \frac{1}{\p_0} \bigg[
        \frac{\1(\Gamma = 1)
        (1-\pi_{\Gamma,0})}{\pi_{\Gamma,0}}
        \left( \frac{\1(A = 1)}{\pi_{A,0}}
        - \frac{\1(A = 0)}{1 -\pi_{A,0}}
        \right) \D(t) \notag \\
        &\quad + \1(\Gamma = 0)
        \left( S_0^*(t)
        - \theta_{T, \Po}(t) \right) \bigg],
    \end{align}
    \begin{equation}\label{eq:gen_eif}
        \varphi_{\theta_G}(O;\eta_0, t)
        = \frac{\1(\Gamma = 1)}
        {\pi_{\Gamma,0}}
        \left( \frac{\1(A = 1)}{\pi_{A,0}}
        - \frac{\1(A = 0)}{1 -\pi_{A,0}}
        \right) \D(t)
        + \left( S_0^*(t)
        - \theta_{G, \Po}(t) \right).
    \end{equation}

    Under the additional
    Assumptions~\ref{ass:param}
    through~\ref{ass:hem}, if there
    exists $\epsilon^* > 0$ such that \newline
    $\min\{\pi_{A,0}, \pi^*_{\Gamma,0},
    \Go(m)\} \geq \epsilon^*$ for all
    $m = 1, \dots, K$,
    $\Po$-almost everywhere with \newline
    $\So(t \mid a, w, \gamma) > 0$, then
    $\lambda_{T,\Po}(t)$ and
    $\lambda_{G,\Po}(t)$ are pathwise
    differentiable in $\Mset$ with efficient
    influence functions
    \begin{align}\label{eq:trans_eff_eif}
        \varphi_{\lambda_T}(O;\eta^*_0, t)
        &= \frac{1}{\p_0} \bigg[
        \frac{\1(\Gamma = 1)
        (1-\pi^*_{\Gamma,0})}
        {\pi^*_{\Gamma,0}}
        \left( \frac{\1(A = 1)}{\pi_{A,0}}
        - \frac{\1(A = 0)}{1 -\pi_{A,0}}
        \right) \D(t)
        \notag \\
        &\quad + (1-\pi^*_{\Gamma,0})
        \left( \fo(t)
        - \Eo\left[\fo(t) \mid Z\right]
        \right)
        + \1(\Gamma = 0)
        \left( \Eo\left[\fo(t) \mid Z\right]
        - \lambda_{T, \Po}(t) \right) \bigg],
    \end{align}
    \begin{equation}\label{eq:gen_eff_eif}
        \varphi_{\lambda_G}(O;\eta^*_0, t)
        = \frac{\1(\Gamma = 1)}
        {\pi^*_{\Gamma,0}}
        \left( \frac{\1(A = 1)}{\pi_{A,0}}
        - \frac{\1(A = 0)}{1 -\pi_{A,0}}
        \right) \D(t)
        + \left( \fo(t)
        - \lambda_{G, \Po}(t) \right).
    \end{equation}
\end{theorem}

Each EIF consists of two components.
The first is an inverse probability
weighted (IPW) correction involving
the term $\D(t)$, which adjusts
for the weighted residual between the hazard
$h_0(m)$ and the observed event indicator
$\1(L_m = 1)$ at each time point,
weighted by the inverse of the at-risk
probability $\So(m) \Go(m)$.
This term closely parallels the
augmentation terms found in classical
causal survival inference
\citep{diaz2019statistical,
luedtke2017sequential}.
The second component is an outcome
regression augmentation that corrects the
plug-in estimate, where
$\mu \in \{\theta_T, \theta_G,
\lambda_T, \lambda_G\}$ denotes
a generic estimand: $S^*_0(t) - \mu(t)$
for the base estimators
and $\fo(t,V) - \mu(t)$ for the
structured estimators.
The interplay between these two
components produces the double robustness
property established in
Theorem~\ref{thm:drc} below.

The transport EIFs
$\varphi_{\theta_T}$ and
$\varphi_{\lambda_T}$ restrict the
$\D$ correction to source subjects
$(\Gamma = 1)$ and reweight to the
target population through the odds ratio
$(1 - \pi_{\Gamma,0}) / \pi_{\Gamma,0}$
(or its $Z$-conditional analog
for $\lambda_T$).
Since outcome data are observed only
in the source population, the
generalization EIFs
$\varphi_{\theta_G}$ and
$\varphi_{\lambda_G}$ also evaluate
$\D$ on source subjects only,
with the population propensity
$1/\pi_{\Gamma,0}$ (or $1/\pi^*_{\Gamma,0}$)
serving to reweight from the source
to the combined population.
Target subjects contribute through the
outcome regression augmentation alone.

A key structural difference between the
$\theta$'s and $\lambda$'s EIFs is the propensity
score used for population reweighting.
The base estimators use
$\pi_{\Gamma,0}$,
which conditions on the full covariate
vector.
The structured estimators replace this with
$\pi^*_{\Gamma,0}$,
conditioning only on the
lower-dimensional subset $Z$.
Since $Z \subseteq V \subseteq W$, we expect
the ratio $\frac{1-\pi^*_{\Gamma}}{\pi^*_{\Gamma}}$
to be more stable than $\frac{1-\pi_{\Gamma}}{\pi_{\Gamma}}$,
and thus the $\lambda$ estimators having lower variance
and better efficiency than the $\theta$ estimators \citep{Bickel97}.
However, if the correlation between $\pi^*_{\Gamma}$ and the outcome
is smaller than the correlation between $\pi_{\Gamma}$ and the outcome
this efficiency gain may not hold \citep{rudolph2025transporting}.
In the majority of applied cases we expect a reduction of variance.


\subsection{Von Mises expansion and
second-order remainder}

The theoretical properties of the
proposed estimators are established
through the von Mises expansion
of each estimand functional.
For a functional $\Psi(\P)$
and a candidate influence function
$\varphi(O; \P)$, the von Mises
expansion takes the form
\[
\Psi(\hat \P) - \Psi(\Po) =
-\Po \varphi_{\hat \P} +
\R_2(\hat \P, \Po),
\]
where $\R_2$ is the second-order
remainder.
The key property of the EIF is that
$\R_2$ consists entirely of products
of differences between true and estimated
nuisance parameters, so that
$\R_2 = 0$ when any one ``factor''
is evaluated at the truth.
This product structure is what enables
both the double robustness and the
mixed-bias rate conditions for
asymptotic normality.

\begin{theorem}{Von Mises expansion
remainder}\label{thm:vonmises}
    The second-order remainder at
    restriction time $\tau$ of the
    von Mises expansion for the base
    estimators satisfies
    \begin{equation}\label{eq:R2_base}
        \R_2(\hat \eta; \theta) =
        \sum_{m = 1}^{\tau}
        \sum_{k=0}^{m-1}
        O_{\hat \P} \left[
        \| h_0(m) - \hat h(m) \|
        \left( \|g_{c,0}(k) - \hat g_C(k)\|
        + \| \pi_{A,0} - \hat \pi_A \|
        + \| \pi_{\Gamma,0}
        - \hat \pi_{\Gamma} \|
        \right) \right],
    \end{equation}
    for $\theta_T(\tau)$ and
    $\theta_G(\tau)$.
    For the structured estimators,
    \begin{align}\label{eq:R2_eff}
    \R_2(\hat \eta^*; \lambda)
    &= \sum_{m = 1}^{\tau}
    \sum_{k=0}^{m-1}
    O_{\hat \P} \bigg[
    \left( \|\fo(\tau) - \hat \f(\tau)\|
    + \| h_0(m) - \hat h(m) \| \right)
    \notag \\
    &\quad \times \left(
    \|g_{c,0}(k) - \hat g_C(k)\|
    + \| \pi_{A,0} - \hat \pi_A \|
    + \| \pi^*_{\Gamma,0}
    - \hat \pi^*_{\Gamma} \|
    \right) \bigg],
    \end{align}
    for $\lambda_T(\tau)$ and
    $\lambda_G(\tau)$.
\end{theorem}

The remainder (\ref{eq:R2_base}) for
the base estimators is a sum over
time points of products between the
hazard estimation error $\|h_0 - \hat h\|$
and the combined errors in the censoring
hazard and propensity scores.
For the structured estimators,
(\ref{eq:R2_eff}) includes the additional
term $\|\fo - \hat \f\|$ in the first
factor, reflecting the extra nuisance
introduced by the additive parameterization.
The contrast $\fo$ and the hazard $h_0$
are not independent nuisances, since
$\fo$ is a smooth functional of $h_0$
under Assumption~\ref{ass:param}
and inherits its estimation error.
Grouping the two terms in the same factor
of the remainder is therefore natural,
and a consistent estimator of $h_0$
yields a consistent estimator of $\fo$.
In both cases, the product structure ensures
that $\R_2 = o_\P(1)$ whenever at least
one factor converges to zero,
which is the basis for double robustness.
For asymptotic normality, the stronger
condition $\R_2 = o_\P(n^{-1/2})$ is
required, which can be achieved if both
factors converge at rate $o_\P(n^{-1/4})$.


\subsection{Estimation algorithm and
asymptotic properties}

We propose a one-step debiased estimator
for each estimand
$\mu \in \{\theta_T, \theta_G,
\lambda_T, \lambda_G\}$,
constructed by subtracting the
leading bias from the plug-in:
$\tilde \mu(t) = \hat \mu(t)
+ n^{-1} \sum_{i=1}^n
\varphi_{\mu}(O_i; \hat \eta)$.
To bypass Donsker class requirements
and allow the use of flexible
data-adaptive regression methods,
we employ sample splitting via
cross-fitting
\citep{klaassen1987consistent,
zheng2011cross, chernozhukov2018double}.

Let $\mathcal{V}_1, \dots, \mathcal{V}_B$
denote a random partition of
$\{1, \dots, n\}$ into $B$ prediction
sets of approximately equal size,
with cardinalities $n_1, \dots, n_B$.
For each fold $b$, define the training
set $\mathcal{V}_{n,b}^c =
\{O_i : i \notin \mathcal{V}_{n,b}\}$
and let
$\hat \pi_{A,n,b}$,
$\hat \pi_{\Gamma,n,b}$,
$\hat \pi^*_{\Gamma,n,b}$,
$\hat h_{n,b}$,
$\hat g_{C,n,b}$, and
$\hat \f_{n,b}$
denote the nuisance estimates obtained
from the training set.
Write $\varphi_{\mu}(O; \hat \eta_{n,b})$
for the EIF with estimated nuisances
substituted for their true values.
The cross-fitted one-step estimator is
computed as follows.

\begin{enumerate}
    \item Estimate propensity scores:
    $\hat \pi_{A,n,b}$ by regressing $A$
    on $W$ in the source population
    ($\Gamma = 1$) and predicting for
    all observations;
    $\hat \pi_{\Gamma,n,b}$ by regressing
    $\Gamma$ on $W$; and
    $\hat \pi^*_{\Gamma,n,b}$ by regressing
    $\Gamma$ on $Z$.
    \item For each $t = 1, \dots, \tau$,
    estimate the conditional hazards
    in the source population:
    \begin{enumerate}
        \item Fit
        $\hat h_{n,b}(t)$
        by regressing $L_t$ on $(A, W)$
        among at-risk source subjects
        ($I_t = 1$, $\Gamma = 1$),
        and predict for all observations.
        Set $\hat S_{n,b}(t) =
        \prod_{m=1}^t
        (1 - \hat h_{n,b}(m))$.
        \item Fit
        $\hat g_{C,n,b}(t)$
        by regressing $R_t$ on $(A, W)$
        among at-risk source subjects
        ($J_t = 1$, $\Gamma = 1$),
        and predict for all observations.
        Set $\hat G_{n,b}(t) =
        \prod_{m=0}^{t-1}
        (1 - \hat g_{C,n,b}(m))$.
    \end{enumerate}
    \item For estimators $\lambda$'s only: Estimate
    $\hat \f_{n,b}(\tau)$ by regressing
    $\hat S_{n,b}(\tau \mid A=1) -
    \hat S_{n,b}(\tau \mid A=0)$
    on $V$ in the source population,
    then predict for all observations.
    \item For estimators $\lambda$'s only: Regress
    $\hat \f_{n,b}(\tau)$ on $Z$ in
    the combined population to obtain
    $\hat \E[\hat \f_{n,b}(\tau) \mid Z]$.
    \item Compute the cross-fitted
    one-step estimator:
    \[\tilde \mu(\tau) = \sum_{b=1}^B
    (n_b / n) \left[\hat \mu_{n,b}(\tau)
    + n_b^{-1} \sum_{i \in \mathcal{V}_{n,b}}
    \varphi_{\mu}(O_i;
    \hat \eta_{n,b}) \right].\]
\end{enumerate}

The variance of $\tilde \mu$ is estimated
as the sample variance of
$\varphi_{\mu}(O_i; \hat \eta_{n,b})$
across all observations.
Confidence intervals and hypothesis
tests follow from the Wald construction
using the asymptotic normality
established in Theorem~\ref{thm:wc}.

\begin{theorem}{Weak convergence}
\label{thm:wc}

    Assume that
    $\R_2(\hat \eta; \theta) =
    o_\P(n^{-1/2})$ and that
    $\hat \pi_A$, $\hat \pi_{\Gamma}$,
    $\hat \S(m)$, and $\hat \G(m)$
    are bounded away from zero
    in probability for all
    $m = 1, \dots, K$. Then
    \[
        \sqrt{n}(\tilde \theta_{T,n}
        - \theta_{T,\Po})
        \overset{d}{\to}
        \mathcal{N}(0,
        \textnormal{Var}[
        \varphi_{\theta_T}(O; \eta_0)]),
        \quad
        \sqrt{n}(\tilde \theta_{G,n}
        - \theta_{G,\Po})
        \overset{d}{\to}
        \mathcal{N}(0,
        \textnormal{Var}[
        \varphi_{\theta_G}(O; \eta_0)]).
    \]
    If additionally
    $\R_2(\hat \eta^*; \lambda) =
    o_\P(n^{-1/2})$ and
    $\hat \pi^*_{\Gamma}$ is bounded
    away from zero in probability, then
    \[
        \sqrt{n}(\tilde \lambda_{T,n}
        - \lambda_{T,\Po})
        \overset{d}{\to}
        \mathcal{N}(0,
        \textnormal{Var}[
        \varphi_{\lambda_T}(O; \eta^*_0)]),
        \quad
        \sqrt{n}(\tilde \lambda_{G,n}
        - \lambda_{G,\Po})
        \overset{d}{\to}
        \mathcal{N}(0,
        \textnormal{Var}[
        \varphi_{\lambda_G}(O; \eta^*_0)]).
    \]
\end{theorem}

\begin{theorem}{Doubly robust consistency}
\label{thm:drc}

    For the base estimators, assume that
    for all $m = 1, \dots, K$ either
    $\|h_0(m) - \hat h(m)\| = o_\P(1)$,
    or for all $k = 0, \dots, K$,
    $\|g_{C,0}(k) - \hat g_C(k)\|
    + \|\pi_{A,0} - \hat \pi_A\|
    + \|\pi_{\Gamma,0}
    - \hat \pi_{\Gamma}\| = o_\P(1)$.
    Then
    $\tilde \theta_T
    = \theta_{T,\Po} + o_\P(1)$ and
    $\tilde \theta_G
    = \theta_{G,\Po} + o_\P(1)$.

    For the structured estimators, assume
    that either
    $\|\fo(\tau) - \hat \f(\tau)\|
    + \|h_0(m) - \hat h(m)\| = o_\P(1)$,
    or
    $\|g_{C,0}(k) - \hat g_C(k)\|
    + \|\pi_{A,0} - \hat \pi_A\|
    + \|\pi^*_{\Gamma,0}
    - \hat \pi^*_{\Gamma}\| = o_\P(1)$.
    Then
    $\tilde \lambda_T
    = \lambda_{T,\Po} + o_\P(1)$ and
    $\tilde \lambda_G
    = \lambda_{G,\Po} + o_\P(1)$.
\end{theorem}

Theorem~\ref{thm:wc} establishes that
all four estimators are asymptotically
normal and semiparametrically efficient
when the second-order remainder
$\R_2 = o_\P(n^{-1/2})$.
By the product structure of
(\ref{eq:R2_base}) and (\ref{eq:R2_eff}),
this condition is satisfied whenever
each factor of the remainder converges
at rate $o_\P(n^{-1/4})$, a rate
achievable by many data-adaptive
regression methods including
$\ell_1$-regularized regression,
random forests, neural networks,
the highly adaptive lasso, and
ensemble learners such as SuperLearner
\citep{bickel2009simultaneous,
wager2015adaptive, chen1999improved,
benkeser2016highly, vdl2007super}.

Theorem~\ref{thm:drc} provides a weaker
guarantee: consistency holds whenever
at least one factor of the remainder
converges to zero at any rate.
This double robustness property protects
against misspecification of either the
outcome model (hazard $h$ and contrast $\f$)
or the weighting model
(propensity scores and censoring hazard).
The conditions for double robust
consistency are strictly weaker than
those for asymptotic normality.
Consistency does not guarantee
valid inference, as the convergence
rate may be slower than $n^{-1/2}$.
This distinction is explored empirically
in Section~\ref{sec:num}.
All proofs are given in
Web Appendix~B.


\section{Numerical Study}\label{sec:num}

\subsection{Data-generating mechanism}
\label{sec:num_DMG}

For each simulation replicate, we generate
a sample of size $n$ from a discrete-time
survival model observed over
$K = 5$ follow-up times
$t \in \{1, \ldots, 5\}$.
The baseline covariate vector is
$W = (W_1, W_2, W_3, W_4, W_5)$,
with each component drawn independently
from a $\mathrm{Uniform}(0,1)$ distribution.

Population membership
$\Gamma \in \{0,1\}$ is generated
from the logistic model
\[
\P(\Gamma = 1 \mid W) =
\mathrm{expit}(0.5 + W_3 - 2 W_1),
\]
so that larger values of $W_3$ and
smaller values of $W_1$ increase
the probability of belonging to the
source population.
The covariates whose distributions
differ across populations are
$X = (W_1, W_3)$, while
$W_2$, $W_4$, and $W_5$ retain the
same marginal distributions in both
populations.
Treatment $A \in \{0,1\}$ is assigned
using different mechanisms in
the two populations:
\[
\P(A = 1 \mid W, \Gamma) =
\begin{cases}
    \mathrm{expit}(-1 + W_3
    + 0.6\, W_1 + 0.4\, W_2),
    & \Gamma = 0,\\[4pt]
    0.5, & \Gamma = 1.
\end{cases}
\]
In the source population ($\Gamma = 1$),
treatment is randomized with
probability $1/2$.
In the target population ($\Gamma = 0$),
treatment depends on $(W_1, W_2, W_3)$,
introducing confounding.

Generating discrete-time survival data
that simultaneously conform to
Assumptions~\ref{ass:condex}
through~\ref{ass:hem} is nontrivial,
as directly specifying the hazard can
violate monotonicity of the survival
function or reintroduce dependence
on non-effect-modifying covariates.
We adopt a constructive four-step
procedure (detailed in
Web Appendix~C).
We first specify temporary
treatment-specific hazards
\[
h^{\mathrm{temp}}(t \mid W, A) =
\mathrm{expit}(-1.5 + 0.1\,t - 0.5\, W_1
+ A \times (- W_2 - W_3 + W_2 W_3)),
\]
and their corresponding survival functions
$S_a^{\mathrm{temp}}(t \mid W) =
\prod_{m=1}^t
\{1 - h^{\mathrm{temp}}_a(m \mid W)\}$.
The preliminary treatment contrast is
$\f_{\mathrm{temp}}(t,W) =
S_1^{\mathrm{temp}}(t \mid W)
- S_0^{\mathrm{temp}}(t \mid W)$.

To satisfy Assumption~\ref{ass:tem}, we let
$V = (W_2, W_3)$ be the true
effect modifier subset, so that the
outcome difference depends on the covariates only through $V$.
The common subset driving both
covariate shift and effect modification
is $Z = X \cap V = \{W_3\}$,
satisfying the partial
study-heterogeneity structure of
Assumption~\ref{ass:hem}.
To enforce Assumption~\ref{ass:tem},
we project $\f_{\mathrm{temp}}(t, W)$
onto the $\sigma(V)$-field:
\[
\f(t, V) =
\mathbb{E}[\f_{\mathrm{temp}}(t, W)
\mid W_2, W_3],
\]
approximated via flexible regression.
The final survival functions are
reconstructed as
$S_0(t \mid W) = S_0^{\mathrm{temp}}
(t \mid W)$ and
$S_1(t \mid W) =
S_0^{\mathrm{temp}}
(t \mid W) + \f(t,V)$.
The final hazards are obtained via
$h_a(t \mid W) = 1 - S_a(t \mid W)
/ S_a(t-1 \mid W)$
with $S_a(0 \mid W) = 1$.
Censoring is generated through
\[
g_C(t \mid A, W, \Gamma) =
\mathrm{expit}(-5 + 0.20\,t
+ 0.20\,W_1 - 0.30\,W_3
- \Gamma + 0.15\,A).
\]


\subsection{Simulation design}
\label{sec:num_design}

To evaluate the finite-sample properties
of the proposed estimators and to
empirically assess the asymptotic
double robustness and normality results
from Section~\ref{sec:properties},
we conducted Monte Carlo simulations
under three nuisance estimation scenarios.
To mimic a real applied setting, 
all nuisance functions were estimated
using the data-adaptive ensemble learner
SuperLearner \citep{vdl2007super}.
Under flexible specification, the
SuperLearner library consisted of
generalized linear models
(\texttt{glm}),
generalized linear models with
interactions (\texttt{glm.interaction}),
random forests (\texttt{ranger}),
and multivariate adaptive regression
splines (\texttt{earth}).
All computations were carried out in
R version 4.3.0 \citep{Rcore2023}. The
SuperLearner library was implemented with
the \texttt{SuperLearner} (version 2.0-28)
\citep{superlearnerR}, \texttt{ranger}
(version 0.15.1) \citep{rangerR}, and
\texttt{earth} (version 5.3.2)
\citep{earthR} packages from CRAN.
To induce misspecification, the relevant
nuisance parameters were estimated
using only the empirical mean learner
(\texttt{mean}).
The three scenarios are:
\begin{enumerate}
    \item \textbf{Flexible specification.}
    All nuisance functions are estimated
    with the full SuperLearner library.
    \item \textbf{Misspecified hazard.}
    The survival hazard $\hat h$ is
    estimated using only the empirical
    mean learner, while propensity scores
    and censoring hazard use the full library.
    \item \textbf{Misspecified propensities
    and censoring.}
    The propensity scores
    $\hat \pi_A$, $\hat \pi_{\Gamma}$
    and the censoring hazard $\hat g_C$
    are estimated using only the empirical
    mean learner, while the survival hazard
    uses the full library.
\end{enumerate}
Scenarios 2 and 3 correspond to the
two branches of the double robustness
condition in Theorem~\ref{thm:drc}:
under correct specification of the
remaining nuisance parameters, the
estimators should remain consistent
despite misspecification of the other set.

For each scenario, we generated
$S = 100$ independent datasets for
each sample size
$n \in \{500, 1000, 2000, 3000, 4000\}$.
The true estimand values were computed
by Monte Carlo averaging over a single
large reference dataset of size
$n_{\mathrm{ref}} = 1{,}000{,}000$.
For each replicate and time point $t$,
we computed all four estimators and
evaluated the following performance
metrics:
\begin{enumerate}[label=(\roman*)]
    \item Bias:
    $S^{-1}\sum_{s=1}^{S}
    [\tilde\mu_{n,s}(t) - \mu_{\Po}(t)]$.
    \item Scaled bias:
    $\sqrt{n} \cdot S^{-1}\sum_{s=1}^{S}
    [\tilde\mu_{n,s}(t) - \mu_{\Po}(t)]$.
    \item Scaled mean squared error:
    $n \cdot S^{-1}\sum_{s=1}^{S}
    [\tilde\mu_{n,s}(t) - \mu_{\Po}(t)]^2$.
    \item Coverage probability:
    $S^{-1}\sum_{s=1}^{S}
    \1\bigl(\mu_{\Po}(t) \in
    [\tilde\mu^L_{n,s}(t),\,
    \tilde\mu^H_{n,s}(t)]\bigr)$,
\end{enumerate}
where $\tilde\mu^L_{n,s}(t)$ and
$\tilde\mu^H_{n,s}(t)$ are the bounds
of the $95\%$ Wald confidence interval
from replicate $s$, and
$\mu \in \{\theta_T, \theta_G,
\lambda_T, \lambda_G\}$ denotes
any of the four estimators.
Under flexible specification, the
scaled bias should stabilize as $n$
increases (indicating $\sqrt{n}$
consistency), the scaled MSE should
converge to the semiparametric
efficiency bound, and coverage
should approach the nominal $95\%$ level.
Under misspecification, double robustness
predicts that the unscaled bias should
decrease with $n$, though the rate
may be slower than $n^{-1/2}$.


\subsection{Results}\label{sec:num_res}

\begin{figure}[!ht]
\centering
\includegraphics[width=\textwidth,height=0.82\textheight,keepaspectratio,%
alt={Grid of five panels summarizing integrated simulation performance
of the four estimators (theta_T, theta_G, lambda_T, lambda_G) over
sample sizes 500 to 4000 under three nuisance estimation scenarios
(flexible specification, misspecified hazard, misspecified propensities
and censoring). Panels show integrated bias, integrated scaled bias,
integrated scaled mean squared error, integrated variance, and mean
coverage probability across time points t = 1 to 5.}]{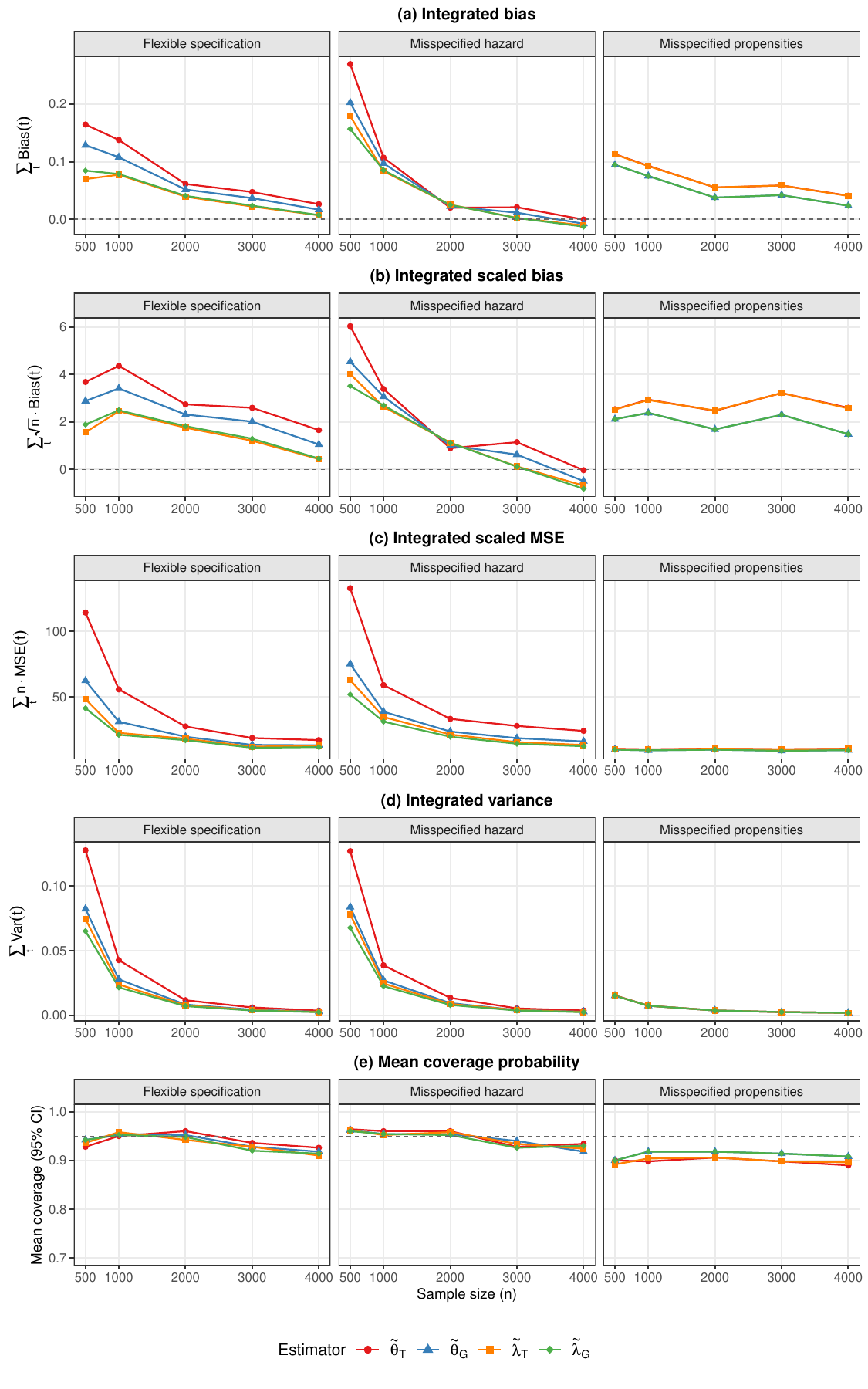}
\caption{Integrated simulation performance
metrics across time points
$t = 1, \ldots, 5$ for the four
proposed estimators under three
nuisance estimation scenarios.
Panels show (a) integrated bias,
(b) integrated scaled bias,
(c) integrated scaled MSE,
(d) integrated variance, and
(e) mean coverage probability.
Results are based on $S = 100$
replicates at each sample size.}
\label{fig:sim_results}
\end{figure}

Figure~\ref{fig:sim_results} presents the
integrated simulation performance metrics
across all time points for the three
nuisance estimation scenarios.
Table~\ref{tab:sim_results} reports the
corresponding numerical values.

Under flexible specification
(Figure~\ref{fig:sim_results}, left column),
the integrated bias decreases with $n$
for all four estimators, and the scaled
bias stabilizes, consistent with
$\sqrt{n}$ consistency.
The scaled MSE converges toward
the semiparametric efficiency bound,
and coverage approaches the nominal
$95\%$ level at larger sample sizes.
The structured estimators
$\tilde \lambda_T$ and
$\tilde \lambda_G$ achieve
lower integrated variance and scaled
MSE than their base counterparts
$\tilde \theta_T$ and
$\tilde \theta_G$, confirming
the theoretical efficiency gains
from leveraging the known effect
modifier subset.
This gain is most pronounced at
smaller sample sizes, where the
variance reduction from reweighting
over $Z$ rather than $W$ has the
largest practical impact.
The relative efficiency column in
Table~\ref{tab:sim_results} quantifies
these gains: $\tilde\lambda_T$ achieves
a 1.7-fold variance reduction over
$\tilde\theta_T$ at $n = 500$,
declining smoothly to a 1.4-fold
reduction at $n = 4000$, with smaller
but consistent gains of 1.1 to 1.3-fold
for $\tilde\lambda_G$ over $\tilde\theta_G$.
The larger gain for the transport pair
is consistent with transport reweighting
being more sensitive to the dimensionality
of the population propensity score than
the generalization reweighting.

Under misspecified hazard
(Figure~\ref{fig:sim_results},
center column), the bias decreases
with $n$ despite the outcome model
being estimated with only the
empirical mean learner.
This demonstrates the double
robustness property established in
Theorem~\ref{thm:drc}: the estimators
remain consistent when the propensity
scores and censoring hazard are
correctly specified, even if the
survival hazard is severely misspecified.
At $n = 4000$, the bias is near zero
for all estimators.
The convergence is slower than under
flexible specification, as expected
since the rate depends on the product
of nuisance estimation errors
(Theorem~\ref{thm:vonmises}).

Under misspecified propensities
and censoring
(Figure~\ref{fig:sim_results},
right column), the estimators again
exhibit decreasing bias with $n$,
confirming double robustness from the
complementary direction: the correctly
specified survival hazard ensures
consistency even when the propensity
scores and censoring model are
misspecified.
In this scenario, the base and structured
estimators perform nearly identically,
since the misspecified propensity scores
(estimated as constants by the empirical
mean learner) eliminate the differential
weighting between
$\pi_{\Gamma,0}(W)$ and
$\pi^*_{\Gamma,0}(Z)$ that drives the
efficiency gain.
The coverage in this scenario is
below the nominal $95\%$ level,
reflecting the fact that double
robust consistency does not guarantee
the convergence rate required for
valid asymptotic inference
(Theorem~\ref{thm:wc}).


\section{Illustrative Application:
Hormone Therapy and Coronary Heart
Disease in the WHI}

The relationship between postmenopausal
hormone therapy (HT) and coronary
heart disease (CHD) has been one of
the most consequential debates in
modern epidemiology, and serves as a
canonical example of the challenges
that motivate formal transportability
and generalizability methods.
For decades, large observational studies
reported substantial reductions in CHD
risk among HT users.
The Nurses' Health Study, among the
largest and longest-running prospective
cohorts, found a 40--50\% reduction
in coronary events associated with
estrogen use
\citep{stampfer1991postmenopausal,
grodstein2000prospective}.
These findings shaped clinical guidelines
and led to widespread HT prescribing
for cardiovascular protection.

The first experimental challenge came
from the Heart and Estrogen/Progestin
Replacement Study (HERS), a randomized
secondary prevention trial that found
no cardiovascular benefit and evidence
of early harm
\citep{hulley1998randomized}.
The Women's Health Initiative (WHI)
estrogen-plus-progestin trial then
confirmed these findings in a primary
prevention setting, documenting
a significant elevation in CHD risk
among women randomized to HT
\citep{rossouw2002risks}.
The estrogen-alone arm yielded a more
nuanced result, with no significant
overall CHD effect
\citep{anderson2004effects}.
The contradiction between observational
and experimental findings prompted
extensive investigation into confounding,
healthy-user bias, and treatment effect
heterogeneity as possible explanations
\citep{prentice2005combined,
prentice2006combined,
hernan2008observational,
manson2013menopausal}.
A central hypothesis is that
the populations enrolled in the
trial and those studied observationally
differed systematically in covariates
that modify the cardiovascular response
to HT, making the WHI a natural
setting for transportability analysis.

The WHI enrolled over 160{,}000
postmenopausal women into two
harmonized components: a set of
randomized clinical trials and a
large prospective observational
study (OS)
\citep{rossouw2002risks,
anderson2004effects}.
Both components share standardized
baseline assessments, identical
follow-up procedures, and rigorously
adjudicated CHD outcomes based on
blinded physician review and medical
record abstraction.
We designate the clinical trial as
the source population ($\Gamma = 1$)
and the observational study as
the target population ($\Gamma = 0$).
Women in the OS were on average
younger, more physically active,
and of higher socioeconomic status
than trial participants, and their
HT use reflected clinical
decision-making and patient
preference rather than random
assignment
\citep{prentice2006combined}.
These systematic population differences
create the covariate shift structure
that our transport and generalization
estimands are designed to address.

The binary exposure is $A \in \{0,1\}$,
where $A = 1$ denotes assignment to
(or self-reported use of) hormone
therapy and $A = 0$ denotes placebo
or non-use.
The primary outcome is time to
incident CHD, defined as the first
adjudicated myocardial infarction
or CHD death.
The observed data align with
our framework:
$O = (W, A, \Gamma, \Delta, \check{T})$,
where $\check{T} = \min(T, C)$
is the observed follow-up time,
$\Delta$ is the event indicator,
and $W$ comprises a rich set of
baseline covariates measured across
both study components.
Outcome data (event times and
censoring) are available in both the
trial and observational arms of the WHI,
though our methodology requires only
source-population outcome data for
the transport estimators and can
leverage target-population data for
the generalization estimators when
available.

Prior work on the WHI provides strong
clinical guidance for specifying the
effect modifier and population shift
variables required by
Assumptions~\ref{ass:param}
through~\ref{ass:hem}.
Age, years since menopause, smoking
status, and cardiometabolic risk factors
(BMI, hypertension, impaired glucose
tolerance) have been shown to modify
the cardiovascular response to HT
\citep{rossouw2002risks,
manson2013menopausal,
prentice2005combined}.
At the same time, variables such as
age, education, physical activity,
and comorbidity burden differ
substantially between the trial
and observational cohorts
\citep{prentice2006combined}.
We define $V$ as a subset of
clinically established effect modifiers,
$X$ as the covariates exhibiting
distributional differences across
$\Gamma$, and $Z = V \cap X$ as
the effect modifiers whose
distributions differ between
populations.
This specification satisfies the
structural conditions required by
our structured estimators: the additive
parameterization (Assumption~\ref{ass:param}),
the effect modifier structure
(Assumption~\ref{ass:tem}), and
partial study heterogeneity
(Assumption~\ref{ass:hem}).

We compute all four estimators at
restriction time $\tau = 7$.
Table~\ref{tab:results_t7} reports
point estimates, 95\% Wald confidence
intervals, estimated variance,
and standard errors for the base
estimators ($\tilde \theta_G$,
$\tilde \theta_T$) and the structured
estimators ($\tilde \lambda_G$,
$\tilde \lambda_T$) that leverage
the known effect modifier subset.

Across all four estimators, the point
estimates of the survival difference
are close to zero, consistent with
the established finding that HT does
not confer a net CHD benefit at the
population level
\citep{rossouw2002risks,
manson2013menopausal}.
None of the 95\% confidence intervals
exclude zero.
The generalization and transport
estimates agree closely within each
class of estimator, as expected when
the covariate shift between populations
is moderate relative to the sample size.

The central finding of
Table~\ref{tab:results_t7} is
the substantial efficiency gain
achieved by the estimators that
leverage the known effect modifier
subset.
The base generalization estimator
$\tilde \theta_G$ has a standard error
of 1.23, while its structured counterpart
$\tilde \lambda_G$ achieves 0.586,
a reduction of more than 52\%.
In terms of variance, $\tilde \lambda_G$
attains 0.343 compared to 1.52
for $\tilde \theta_G$, a 4.4-fold
reduction.
A parallel improvement holds for
the transport estimators:
$\tilde \lambda_T$ achieves a variance
of 0.350 compared to 2.28 for
$\tilde \theta_T$, a 6.5-fold reduction,
with the standard error decreasing
from 1.51 to 0.592.
These gains follow directly from
the semiparametric theory developed
in Section~\ref{sec:properties}:
the structured estimators replace
the full-dimensional reweighting
over $W$ with reweighting over
the lower-dimensional $Z$, reducing
the variance of the population
membership weights.
The resulting confidence intervals
for the structured estimators are
roughly half the width of those
for the base estimators.

This application demonstrates the
practical value of our methodology
in a setting that originally motivated
much of the transportability literature.
The framework accommodates the complex
covariate distributions and clinically
informed effect modifier structure
present in the WHI data.
The efficiency gains from the additive
parameterization are not merely
theoretical: they translate into
substantially narrower confidence
intervals and more precise treatment
effect estimates, even in a large-scale
epidemiologic study with over
160{,}000 participants.


\section{Discussion}

We developed doubly robust,
semiparametrically efficient estimators
for transporting and generalizing causal
survival differences from a trial to
a target population in discrete time.
The structured versions exploit known
effect modifier subsets to reduce the
dimensionality of the reweighting,
yielding variance reductions that are
both theoretically guaranteed and
empirically substantial, as confirmed
in simulations and in the WHI application.

Several limitations apply.
The transportability assumption
(Assumption~\ref{ass:trans}) is
untestable and requires that the
measured covariates capture all sources
of population heterogeneity relevant
to the outcome.
Limited covariate overlap can inflate
the variance of the transport estimators
through extreme inverse probability
weights, though the structured estimators
partially mitigate this by reweighting
over the lower-dimensional $Z$.
The additive parameterization
(Definition~\ref{ass:param}) and the
correct specification of $V$ and $Z$
(Assumptions~\ref{ass:tem}
and~\ref{ass:hem}) are required only
by the structured estimators; the base
estimators remain valid under the
weaker nonparametric model when these
structural conditions are violated.
Computationally, fitting nuisance models
at each time point within each
cross-fitting fold is demanding when
$K$ is large and ensemble learners
are used.

A natural extension would adapt ideas
from the super-efficiency literature
\citep{benkeser2020nonparametric}
to select the effect modifier subset
from the data while preserving valid
inference, achieving the structured
model's variance bound when the
selection is correct and falling back
to the nonparametric bound otherwise.
Further directions include extension to
continuous-time hazard models and
sensitivity analysis for violations of
the transportability assumption.


\section*{Acknowledgments}

The authors declare no conflicts of interest.

\paragraph{Funding.}
This article is based upon work supported by
the National Science Foundation under Grant No.\ 2306556,
and the National Institute of Health Grant No.\ 1R01AI197146-01.

\paragraph{Author contributions.}
Axel Martin: Conceptualization, Methodology,
Software, Formal analysis, Investigation,
Visualization, Writing.
Iv\'an D\'iaz: Conceptualization, Methodology,
Supervision, Writing.
Michele Santacatterina: Conceptualization,
Methodology, Supervision, Writing.


\section*{Data availability}

The applied example uses data from the
Women's Health Initiative (WHI). These
data are not publicly available without
restriction. They can be obtained from
the Women's Health Initiative
(\url{https://www.whi.org}) and through
the National Heart, Lung, and Blood
Institute Biologic Specimen and Data
Repository Information Coordinating Center
(BioLINCC), subject to the required
approvals and data use agreements. The
authors accessed the data under these
terms and are not permitted to
redistribute them. The simulated data
analyzed in Section~\ref{sec:num} are
fully synthetic and can be regenerated
from the accompanying software. An R
package implementing the proposed
estimators and reproducing all simulation
results, \texttt{transSurv}, is provided
as supplementary material.


\section*{Supplementary Materials}

Web Appendices A through D, referenced in
Sections~\ref{sec:id},
\ref{sec:properties}, and~\ref{sec:num},
are available with this paper at the
Biometrics website on Oxford Academic.
An R package implementing the proposed
estimators is also available as
supplementary material.

\newpage
\bibliographystyle{plainnat}
\bibliography{main}

\begin{table}[!ht]
\centering
\caption{Integrated simulation performance
metrics. Bias, scaled bias, and scaled MSE
are summed over $t = 1, \ldots, \tau$.
Coverage is averaged over time points.
RE is the relative efficiency
$\widehat{\mathrm{Var}}(\tilde\theta)
/ \widehat{\mathrm{Var}}(\tilde\lambda)$
of the structured estimator over its
base counterpart within each pair.
Based on $S = 100$ replicates per
sample sizes $500$, $2000$ and $4000$ for clarity.}
\label{tab:sim_results}
\footnotesize
\begin{tabular}{llrrrrrr}
\toprule
$n$ & Estimator &
Bias & Sc.\ Bias & Sc.\ MSE &
Int.\ Var.\ & Coverage & RE \\
\midrule
\multicolumn{8}{l}{\textbf{Flexible specification}} \\
500  & $\tilde\theta_T$ & 0.165 & 3.68 & 114.3 & 0.128 & 0.928 &      \\
     & $\tilde\theta_G$ & 0.129 & 2.88 & 62.5  & 0.082 & 0.942 &      \\
     & $\tilde\lambda_T$ & 0.070 & 1.56 & 48.3 & 0.075 & 0.936 & 1.71 \\
     & $\tilde\lambda_G$ & 0.084 & 1.89 & 41.4 & 0.065 & 0.942 & 1.26 \\
2000 & $\tilde\theta_T$ & 0.061 & 2.73 & 27.6  & 0.012 & 0.960 &      \\
     & $\tilde\theta_G$ & 0.052 & 2.31 & 19.7  & 0.008 & 0.952 &      \\
     & $\tilde\lambda_T$ & 0.039 & 1.75 & 18.2 & 0.008 & 0.942 & 1.51 \\
     & $\tilde\lambda_G$ & 0.041 & 1.82 & 17.1 & 0.007 & 0.948 & 1.17 \\
4000 & $\tilde\theta_T$ & 0.026 & 1.65 & 17.2  & 0.004 & 0.926 &      \\
     & $\tilde\theta_G$ & 0.017 & 1.05 & 13.2  & 0.003 & 0.918 &      \\
     & $\tilde\lambda_T$ & 0.007 & 0.43 & 12.9 & 0.003 & 0.910 & 1.44 \\
     & $\tilde\lambda_G$ & 0.007 & 0.46 & 11.8 & 0.002 & 0.914 & 1.08 \\
\midrule
\multicolumn{8}{l}{\textbf{Misspecified hazard}} \\
500  & $\tilde\theta_T$ & 0.270 & 6.03 & 132.9 & 0.127 & 0.964 &      \\
     & $\tilde\theta_G$ & 0.203 & 4.53 & 75.1  & 0.084 & 0.962 &      \\
     & $\tilde\lambda_T$ & 0.179 & 4.01 & 62.9 & 0.078 & 0.962 & 1.62 \\
     & $\tilde\lambda_G$ & 0.157 & 3.51 & 51.9 & 0.068 & 0.960 & 1.24 \\
2000 & $\tilde\theta_T$ & 0.020 & 0.89 & 33.4  & 0.014 & 0.960 &      \\
     & $\tilde\theta_G$ & 0.022 & 1.00 & 23.7  & 0.009 & 0.954 &      \\
     & $\tilde\lambda_T$ & 0.025 & 1.12 & 21.4 & 0.009 & 0.958 & 1.55 \\
     & $\tilde\lambda_G$ & 0.025 & 1.11 & 19.8 & 0.008 & 0.952 & 1.19 \\
4000 & $\tilde\theta_T$ & $-$0.001 & $-$0.04 & 24.2  & 0.004 & 0.934 &      \\
     & $\tilde\theta_G$ & $-$0.008 & $-$0.50 & 16.2  & 0.003 & 0.918 &      \\
     & $\tilde\lambda_T$ & $-$0.011 & $-$0.69 & 13.5 & 0.003 & 0.924 & 1.48 \\
     & $\tilde\lambda_G$ & $-$0.013 & $-$0.82 & 12.5 & 0.002 & 0.930 & 1.13 \\
\midrule
\multicolumn{8}{l}{\textbf{Misspecified propensities and censoring}} \\
500  & $\tilde\theta_T$ & 0.113 & 2.53 & 10.5  & 0.015 & 0.900 &      \\
     & $\tilde\theta_G$ & 0.095 & 2.11 & 9.9   & 0.015 & 0.900 &      \\
     & $\tilde\lambda_T$ & 0.113 & 2.52 & 10.4 & 0.015 & 0.892 & 1.01 \\
     & $\tilde\lambda_G$ & 0.095 & 2.11 & 9.9  & 0.015 & 0.900 & 1.00 \\
2000 & $\tilde\theta_T$ & 0.055 & 2.46 & 10.7  & 0.004 & 0.906 &      \\
     & $\tilde\theta_G$ & 0.038 & 1.68 & 9.9   & 0.004 & 0.918 &      \\
     & $\tilde\lambda_T$ & 0.055 & 2.48 & 10.7 & 0.004 & 0.906 & 1.00 \\
     & $\tilde\lambda_G$ & 0.038 & 1.68 & 9.9  & 0.004 & 0.918 & 1.00 \\
4000 & $\tilde\theta_T$ & 0.041 & 2.59 & 10.6  & 0.002 & 0.890 &      \\
     & $\tilde\theta_G$ & 0.023 & 1.48 & 9.5   & 0.002 & 0.908 &      \\
     & $\tilde\lambda_T$ & 0.041 & 2.56 & 10.6 & 0.002 & 0.896 & 1.00 \\
     & $\tilde\lambda_G$ & 0.023 & 1.48 & 9.5  & 0.002 & 0.908 & 1.00 \\
\bottomrule
\end{tabular}
\end{table}

\clearpage

\begin{table}[!ht]
\centering
\caption{Generalization and transport
estimates of the causal survival
difference for hormone therapy on
CHD at restriction time $\tau = 7$
in the Women's Health Initiative.}
\label{tab:results_t7}
\begin{tabular}{lrrrrr}
\toprule
\textbf{Estimator} &
\textbf{Estimate} &
\textbf{CI Low} &
\textbf{CI High} &
\textbf{Variance} &
\textbf{SE} \\
\midrule
$\tilde{\theta}_G$ (generalization)
    &  0.0000374 &  $-$2.42 &  2.42
    &  1.52  & 1.23 \\
$\tilde{\theta}_T$ (transport)
    &  0.00151   &  $-$2.96 &  2.96
    &  2.28  & 1.51 \\
$\tilde{\lambda}_G$
(str.\ generalization)
    & $-$0.00329 &  $-$1.15 &  1.15
    &  0.343 & 0.586 \\
$\tilde{\lambda}_T$
(str.\ transport)
    & $-$0.00312 &  $-$1.16 &  1.16
    &  0.350 & 0.592 \\
\bottomrule
\end{tabular}
\end{table}

\clearpage

\appendix
\newgeometry{margin=0.6in}
\section*{Supplementary Material for
``\titlepaper'' by Martin, D\'iaz,
and Santacatterina}

\section{Web Appendix A: Identification
(Section \ref{sec:id})} \label{sec:supp_id}

\subsection{Proof of Theorem \ref{thm:id}}

    \textbf{Transport estimator:} We have for equation (\ref{eq:transport}):

    Given the transitive properties of the difference operation we will focus on a single treatment branch $A =a$ for the proofs below.

\begin{align*}
    \Po(T_a > t \mid \Gamma = 0) & = \Eo \left[ \Po(T_a > t \mid W, \Gamma = 0) \mid \Gamma = 0 \right] \text{ (Tower Rule)} \\
    & = \Eo \left[ \Po(T_a > t \mid W, \Gamma = 1) \mid \Gamma = 0 \right] \text{ (A\ref{ass:trans})} \\
    & = \Eo \left[ \Po(T_a > t \mid W, A = a, \Gamma = 1) \mid \Gamma = 0 \right] \text{ (A\ref{ass:condex})} \\
    & = \Eo \left[ \Po(T > t \mid W, A = a, \Gamma = 1) \mid \Gamma = 0 \right] \text{ (A\ref{ass:cons} \& A\ref{ass:pos})} \\
    & = \Eo \left[ \So(t \mid W, A = a, \Gamma = 1) \mid \Gamma =0 \right] \text{ By definition} \\
    & = \Eo \left[ \hprod \left( 1 - h_0(t \mid W, A = a, \Gamma = 1) \right) \mid \Gamma =0 \right] \text{ (A\ref{ass:ign})} 
\end{align*}

    \textbf{Generalization estimator:} We have for equation (\ref{eq:gen}):

\begin{align*}
    \Po(T_a > t) & = \Eo \left[  \Po(T_a > t \mid W) \right]  \text{ (Tower Rule)} \\
    & = \Eo \left[  \Po(T_a > t \mid W, \Gamma = 1) \right]  \text{ (A\ref{ass:trans})} \\
    & = \Eo \left[  \Po(T_a > t \mid W, A = a, \Gamma = 1) \right]  \text{ (A\ref{ass:condex})} \\
    & = \Eo \left[  \Po(T > t \mid W, A = a, \Gamma = 1) \right]  \text{ (A\ref{ass:cons} \& A\ref{ass:pos})} \\
    & = \Eo \left[ \So(t \mid W, A = a, \Gamma = 1) \right] \text{ By definition} \\
    & = \Eo \left[ \hprod \left( 1 - h_0(t \mid W, A = a, \Gamma = 1) \right) \right] \text{ (A\ref{ass:ign})}  
\end{align*}

\subsection{Proof of Theorem \ref{thm:efsid}}

\textbf{Structured Transport estimator:} We have from Section \ref{sec:supp_id} above:

\begin{align*}
    \Po(T_1 > t \mid \Gamma = 0)& - \Po(T_0 > t \mid \Gamma = 0)  = \\
    & \Eo \left[ \So(t \mid W, A =1, \Gamma =1) - \So(t \mid W, A =0, \Gamma =1) \mid \Gamma = 0 \right] \\
    & = \Eo \left[ \f_0(t,W) \mid \Gamma =0 \right] \text{ (Def \ref{ass:param})} \\
    & = \Eo \left[ \f_0(t,V) \mid \Gamma =0 \right] \text{ (A\ref{ass:tem})} \\
    & = \Eo \left[  \Eo \left[ \f_0(t,V) \mid \Gamma = 0, Z \right] \mid \Gamma =0 \right] \text{ (A\ref{ass:hem})} \\
    & = \Eo \left[  \Eo \left[ \f_0(t,V) \mid \Gamma = 1, Z \right] \mid \Gamma =0 \right] \\
    & = \Eo \left[  \Eo \left[ \f_0(t,V) \mid Z \right] \mid \Gamma =0 \right]
\end{align*}

    \textbf{Structured Generalization estimator:} 
    
\begin{align*}
    \Po(T_1 > t )& - \Po(T_0 > t )  = \\
    & \Eo \left[ \So(t \mid W, A =1, \Gamma =1) - \So(t \mid W, A =0, \Gamma =1)\right] \\
    & = \Eo \left[ \f_0(t,W)  \right] \text{ (Def \ref{ass:param})} \\
    & = \Eo \left[ \f_0(t,V)  \right] \text{ (A\ref{ass:tem})} 
\end{align*}

\begin{remark}
    Assumption~\ref{ass:hem} is not required for the
    identification of $\lambda_{G,\Po}(t)$, which is
    complete at $\Eo[f_0(t,V)]$ using only
    Assumptions~\ref{ass:condex}--\ref{ass:tem}.
    However, Assumption~\ref{ass:hem} enters the
    derivation of the efficient influence function
    for $\lambda_{G,\Po}(t)$ in Web Appendix~B.
    Specifically, it enables replacing
    the population membership propensity
    $\pi_{\Gamma,0}(\Gamma \mid W)$
    with the lower-dimensional
    $\pi^*_{\Gamma,0}(\Gamma \mid Z)$
    in the IPW component of the EIF,
    which is the source of the efficiency gain
    over the base generalization estimator
    $\theta_{G,\Po}(t)$.
    This parallels the role of
    Assumption~\ref{ass:hem} in the efficient
    transport estimator $\lambda_{T,\Po}(t)$,
    where it enters both the identification
    (through the $\Eo[\fo(t,V) \mid Z]$ projection)
    and the EIF.
\end{remark}

\section{Web Appendix B: Estimator properties
(Section \ref{sec:properties})}
\label{sec:supp_properties}

\subsection{Proof of Theorem \ref{thm:eif}}

    This section will assume that all data are discrete and make use of Gateaux derivatives in an indexed parametric submodel $\epsilon$ in order to find an influence function candidate for each of the estimators. Then we will show that this candidate has a von-Mises expansion remainder pathwise derivative at $\epsilon = 0$ equals 0. Let the indexed parametric submodel to be defined as $\mathbb{P}_{\epsilon}^* (o) = (1-\epsilon) \mathbb{P}_0 (o) + \delta_o \epsilon$. Where $\delta_o$ is the Dirac measure at $O = o$. Since we we assumed that $O$ is discrete we have pmf

    \begin{align*}
    p_{\epsilon}^* (o) & = (1-\epsilon) p_0(o) + \1(O = o) \epsilon \\
    & = (1-\epsilon) p(\check{t}, \delta, a, w,\gamma) + \1(\check{T} = \check{t}, \Delta = \delta, A = a, W = w, \Gamma = \gamma) \epsilon
\end{align*}
First note that all estimators can be written as a difference of estimators for each treatment arms $A \in \{0,1\}$. For $\theta$ we have:

\[
    \theta_{T,\Po} (t) = \E_0[ \So(t \mid A = 1,w,\Gamma = 1) - \So(t \mid A = 0,w,\Gamma = 1) \mid \Gamma = 0] = \theta_{T,\Po}(t,A=1), - \theta_{T,\Po}(t,A=0),
\]
\[
    \theta_{G,\Po} (t) = \E_0[ \So(t \mid A = 1,w,\gamma) - \So(t \mid A = 0,w,\gamma)] =\theta_{G,\Po}(t,A = 1), - \theta_{G,\Po}(t,A = 0).
\]
We will therefore focus, in a first place, on solving the pathwise derivative for any $A = a \in \{0,1\}$, and further use these results as building blocks for the estimators $\lambda$.

\subsubsection{Efficient influence function for transport estimator}\label{sec:supp_trans_eif}

Let's start by taking the derivate $\theta_{T,\Pe}(t,a)$ at $\epsilon = 0$, here we let $\gamma = 1$ unless specified otherwise for ease of notation.

\begin{align} \label{eq:eif_trans_head}
    \ppe \theta_{T,\Pe}(t,a) \ate & = \ppe \E_0[ \Se(t \mid A = a,w,\gamma) \mid \Gamma = 0 ] \ate \notag \\
    & = \ppe \sum_w \left[ \Se(t \mid a,w,\gamma) p_{\epsilon}^* (w \mid \Gamma = 0) \right] \ate \notag \\
    & = \sum_w \left[ \Se(t \mid a,w,\gamma) \ppe p_{\epsilon}^* (w \mid \Gamma = 0) \ate + p_{\epsilon}^* (w \mid \Gamma = 0) \ppe \Se(t \mid a,w,\gamma) \ate \right] \notag \\
    & = \sum_w \left[ S_0(t \mid a,w,\gamma) \ppe p_{\epsilon}^* (w \mid \Gamma = 0) \ate + p_0 (w \mid \Gamma = 0) \ppe \Se(t \mid a,w,\gamma) \ate \right]
\end{align}
Where:

\begin{equation*}
    \ppe p_{\epsilon}^* (w \mid \Gamma = 0) \ate = \frac{\mathbbm{1}(\Gamma = 0)}{\Po (\Gamma = 0)} \left( \mathbbm{1} (W =w) - \Po (w \mid \Gamma =0) \right)
\end{equation*}
So the first part of the equation can be written as:

\begin{align}\label{eq:eif_trans_1}
    \sum_w & \left[ \So(t \mid a,w,\gamma ) \ppe p_{\epsilon}^* (w \mid \Gamma = 0) \ate \right] \notag \\
    & = \sum_w \left[ \So(t \mid a,w,\gamma) \frac{\mathbbm{1}(\Gamma = 0)}{\Po (\Gamma = 0)} \left( \mathbbm{1} (W =w) - \Po (w \mid \Gamma =0) \right) \right] \notag \\
    & = \frac{\1(\Gamma =0) }{\Po (\Gamma = 0)} \sum_w \left[ \1(W = w)  \So(t \mid a,w,\gamma) - \Po(w \mid \Gamma = 0) \So(t \mid a,w,\Gamma =1) \right] \notag \\
    & = \frac{\1(\Gamma =0) }{\Po (\Gamma = 0)} \sum_w \left[ \1(W = w)  \So(t \mid a,w,\gamma) - \theta_{T,\Po}(t,a) \right]
\end{align}
Thus, in order to solve $\ppe \theta_{T,\Pe}(t,a) \ate$ we need to solve $\ppe \Se(t \mid a,w,\gamma) \ate$.

\begin{align}\label{eq:eif_surv_1}
    \ppe \Se(t \mid a,w,\gamma) \ate & = \ppe \left[ \hprod \left( 1 - \he (m \mid a,w,\gamma) \right)  \right] \ate \notag \\
    & = \ppe \left[ \exp \left( \log \left[ \hprod \left( 1 - \he (m \mid a,w,\gamma) \right) \right] \right)  \right] \ate \notag \\
    & = \ppe \left[ \exp \left(  \hsum \left( 1 - \he (m \mid a,w,\gamma) \right) \right)  \right] \ate \notag \\
    & = \exp \left(  \hsum \left( 1 - h_0 (m \mid a,w,\gamma) \right) \right) \times \ppe \left[ \hsum \left( 1 - \he (m \mid a,w,\gamma) \right) \right] \ate \notag \\
    & = \So (t \mid a,w, \gamma) \times \ppe \left[ \hsum \left( 1 - \he (m \mid a,w,\gamma) \right) \right] \ate \notag \\
    & = \So (t \mid a,w, \gamma) \times  \hsum  \left( - \ppe \he (m \mid a,w,\gamma) \ate \right) 
\end{align}
Now using the definition of $h(m \mid a,w,\gamma)$ we get

\begin{align}\label{eq:eif_surv_2}
    \ppe \he (m) \ate & = \ppe \Pe( L_m = 1 \mid I_m = 1, a,w,\gamma)  \ate \notag \\
    & = \frac{\1 (I_m = 1, A =a , W = w, \Gamma=1)}{\Po (I_m = 1, A =a , W = w, \Gamma = 1) } \big[ \1(L_m = 1) \notag \\
    & - \Po (L_m = 1 \mid I_m = 1, A =a , W = w, \Gamma = 1) \big] \notag \\
    & = \frac{\1 (I_m = 1, A =a , W = w, \Gamma=1)}{\Po (I_m = 1 \mid a , w, \Gamma = 1) \Po (A =a, W = w, \Gamma = 1) } \big[ \1(L_m = 1) - h_0 (m \mid a,w,\gamma) \big] \notag \\
    & = \frac{\1 (A =a , W = w, \Gamma=1)}{\Po (A =a, W = w, \Gamma = 1)} \left[ \frac{\1(I_m = 1) [\1(L_m =1) - h_0(m \mid a,w,\gamma)]}{\Po (I_m = 1 \mid a , w, \Gamma = 1)} \right] \notag \\
    & = \frac{\1 (A =a , W = w, \Gamma=1)}{\Po (A =a, W = w, \Gamma = 1)} \left[ \frac{\1(I_m = 1) [\1(L_m =1) - h_0(m \mid a,w,\gamma)]}{\So(m \mid a,w,\gamma) \Go(m \mid a,w,\gamma) } \right]
\end{align}
After noticing that $\Po (I_m = 1 \mid a , w, \gamma) = \So(m \mid a,w,\gamma) \Go(m \mid a,w,\gamma)$. Integrating results from (\ref{eq:eif_trans_1}), (\ref{eq:eif_surv_1}), (\ref{eq:eif_surv_2}) into (\ref{eq:eif_trans_head}) and summing over all $w$'s yields the following treatment specific derivative

\begin{align*}
    \ppe \theta_{T,\Pe}(t,a) \ate & = \frac{1}{\Po(\Gamma = 0)} \bigg[ \frac{\1(A=a,\Gamma =1) \Po(\Gamma =0 \mid W)}{\Po(A =a \mid W,\Gamma =1) \Po(\Gamma =1 \mid W)} \So(t \mid a,w,\gamma) \\
    & \times \hsum \left( \frac{\1(I_m = 1)[ h_0(m \mid a,w,\gamma) - \1(L_m = 1)]}{\So(m\mid a,w,\gamma) \Go(m \mid a,w,\gamma)} \right) \\
    & + \1(\Gamma =0) \left( \So(t \mid a,w,\Gamma = 1) - \theta_{T,\Po}(t,a)  \right)  \bigg]
\end{align*}
Here we can recognize the term $\D(t)$ defined in Theorem~\ref{thm:eif} which we will use from now on to simplify notation, recall

\[
    \D(t) = \hsum \frac{ \So(t) \1(I_m =1 ) (h_0(m) - \1 (L_m = 1)) }{\So(m) \Go(m)}
\]
Finally by combining both treatment branches and some observed $A = a$ we get the candidate influence function:
\begin{align}\label{eq:cand_trans_eif}
    \varphi_{\theta_T}(O;\eta_0, t) = & \ppe \theta_{T,\Pe}(t) \ate = \ppe \theta_{T,\Pe}(t,A=1) \ate - \ppe \theta_{T,\Pe}(t,A=0) \ate \notag \\
    &= \frac{1}{\Po(\Gamma = 0)} \bigg[ \frac{\1(\Gamma = 1) \Po(\Gamma = 0\ \mid W) }{\Po(\Gamma = 1\ \mid W) }  \notag \\
    & \times \left( \frac{\1(A = 1)}{\Po(A = 1 \mid W, \Gamma=1)} - \frac{\1 (A = 0)}{\Po(A = 0 \mid W, \Gamma=1)} \right) \D(t) \notag \\
    &+ \1(\Gamma = 0) \left( \So(t \mid A =1 , w, \Gamma =1) - \So(t \mid A =0 , w, \Gamma =1)  - \theta_{T, \Po}(t) \right) \bigg] \notag \\
    & \text{\textcolor{blue}{Which can be rewritten in a more condensed form:} } \notag \\
    &= \frac{1}{\p_0} \bigg[ \frac{\1(\Gamma = 1) (1-\pi_{\Gamma,0}) }{\pi_{\Gamma,0} }  \left( \frac{\1(A = 1)}{\pi_{A,0}} - \frac{\1 (A = 0)}{1 -\pi_{A,0}} \right) \D(t) \notag \\
        &+ \1(\Gamma = 0) \left( S_0^*(t) - \theta_{T, \Po}(t) \right) \bigg]
\end{align}
The next step is to find the remainder of this candidate influence function in the von-Mises expansion. Recall the definition of the von-Mises expansion

\begin{align*}
    \Psi(\hat{P}) - \Psi(\Po) & = \int \varphi(o;\hat{P}) \dd(\hat{P} - \Po)(o) + R_2(\hat{P},\Po) \\
    & = \int \varphi(o;\hat{P}) \dd\hat{P}(o) - \int \varphi(o;\hat{P}) \dd\Po(z) + R_2(\hat{P},\Po)
\end{align*}
Where $\int \varphi(o;\hat{P}) \dd\hat{P}(o) = 0$ by construction. Thus:

\begin{align*}
    \Psi(\hat{P}) - \Psi(\Po) & = - E_{\Po}[\varphi(o;\hat{P})] + R_2(\hat{P},\Po)
\end{align*}
and 

\begin{align*}
    R_2(\hat{P},\Po) & = \Psi(\hat{P}) - \Psi(\Po) + E_{\Po}[\varphi(o;\hat{P})] 
\end{align*}
Note that, once again we can focus on a single treatment arm $A=a$ before extending to the difference of the treatment arms. Note we have for a single treatment arm we have candidate

\[
\varphi_{\theta_T}(O;\eta_0, t, a) =\frac{1}{\p_0} \bigg[ \frac{\1(\Gamma = 1) (1-\pi_{\Gamma,0}) }{\pi_{\Gamma,0} } \frac{\1(A = a)}{\pi_{A,0}}  \D(t) + \1(\Gamma = 0) \left( \S_0(t) - \theta_{T, \Po}(t,a) \right) \bigg]
\]

Let us first consider the cross-term where nuisance parameters are estimated under $\hat \P$ but the true parameter $\theta_{T,\Po}(t,a) $ remains the final term of the equation above

\[
\varphi_{\theta_T}(O, \theta_{T, \Po};\hat \eta, t, a) =\frac{1}{\hat \p} \bigg[ \frac{\1(\Gamma = 1,A = a) (1-\hat \pi_{\Gamma}) }{\hat \pi_{\Gamma} \hat \pi_{A}} \hat \D(t) + \1(\Gamma = 0) \left( \hat \S(t) - \theta_{T, \Po}(t,a) \right) \bigg]
\]
Using iterated expectation yields:

\begin{align}\label{eq:cross_R}
    \Eo \left[ \varphi_{\theta_T}(O, \theta_{T, \Po};\hat \eta, t, a) \right] & = \Eo \left[ \Eo \left[ \varphi_{\theta_T}(O, \theta_{T, \Po};\hat \eta, t, a) \mid A,W,\Gamma, I_m \right] \right] \notag \\
    & = \Eo \left[ \frac{1}{\hat \p} \left( \frac{\pi_{A,0} \pi_{\Gamma,0} (1-\hat \pi_{\Gamma}) }{\hat \pi_{\Gamma} \hat \pi_{A}} \hat \D^*(t ) + (1-\pi_{\Gamma,0}) \Eo \left[ \left( \hat \S(t) - \theta_{T, \Po}(t,a) \right) \right]  \right) \right] \notag \\
    & = \Eo \left[ \frac{(1-\pi_{\Gamma,0})}{\hat \p} \left( \frac{\pi_{A,0} \pi_{\Gamma,0} (1-\hat \pi_{\Gamma}) }{\hat \pi_{\Gamma} \hat \pi_{A} (1-\pi_{\Gamma,0})} \hat \D^*(t ) + \Eo \left[ \left( \hat \S(t) - \theta_{T, \Po}(t,a) \right) \right] \right) \right]
\end{align}
Where, importantly

\[
\hat \D^*(t ) = \hat \S(t) \hsum \frac{\So(m) \Go(m) (\hat h(m) - h_0(m) )}{\hat \S(m) \hat \G(m)},
\]
and note that using Lemma \ref{supp_lemma:seq} we have

\begin{align*}
    \Eo \left[  \hat \S(t) - \theta_{T, \Po}(t,a)  \right] & = \Eo \left[ \hprod  ( 1 - \hat h(m \mid a,w,\gamma)) - \hprod  ( 1 - h_0(m \mid a,w,\gamma)) \right] \\
    & = \hsum \left( \So(m) (\hat h(m) - h_0(m)) \frac{\hat \S(t)}{\hat \S(m)} \right)
\end{align*}
Which can be factorized with $\hat \D^*$ in equation (\ref{eq:cross_R}) to give

\begin{align}\label{eq:trans_R2_1}
    \Eo \left[ \varphi_{\theta_T}(O, \theta_{T, \Po};\hat \eta, t, a) \right] & = \Eo \bigg[ \frac{(1-\pi_{\Gamma,0})}{\hat \p} \hsum \So(m) (\hat h(m) - h_0(m)) \frac{\hat \S(t)}{\hat \S(m)} \notag \\
    & \times  \left( \frac{\pi_{A,0} \pi_{\Gamma,0} (1-\hat \pi_{\Gamma}) \Go(m) }{\hat \pi_{\Gamma} \hat \pi_{A} (1-\pi_{\Gamma,0}) \hat \G(m)} -1 \right)  \bigg] \notag \\
    & = \Eo \bigg[ \frac{(1-\pi_{\Gamma,0})}{\hat \p} \hsum \So(m) (\hat h(m) - h_0(m)) \frac{\hat \S(t)}{\hat \S(m)} \notag \\
    & \times  \left( \frac{\pi_{A,0} \pi_{\Gamma,0} (1-\hat \pi_{\Gamma}) (\Go(m) - \hat \G(m)) }{\hat \pi_{\Gamma} \hat \pi_{A} (1-\pi_{\Gamma,0}) \hat \G(m)} - \frac{\pi_{A,0} \pi_{\Gamma,0} (1-\hat \pi_{\Gamma}) - \hat \pi_{\Gamma} \hat \pi_{A} (1-\pi_{\Gamma,0})}{\hat \pi_{\Gamma} \hat \pi_{A} (1-\pi_{\Gamma,0})} \right)  \bigg] \notag \\
    & \text{\textcolor{blue}{Applying Lemma \ref{supp_lemma:seq} on $\Go(m) - \hat \G(m)$} } \notag \\
    & = \Eo \bigg[ \frac{(1-\pi_{\Gamma,0})}{\hat \p} \hsum \So(m) (\hat h(m) - h_0(m)) \frac{\hat \S(t)}{\hat \S(m)} \notag \\
    & \times  \bigg( \frac{\pi_{A,0} \pi_{\Gamma,0} (1-\hat \pi_{\Gamma}) }{\hat \pi_{\Gamma} \hat \pi_{A} (1-\pi_{\Gamma,0}) \hat \G(m)} \left[ \sum_{k = 0}^{m -1} \left( G(k) (\hat g_{C}(k) - g_{C,0}(k)) \frac{\Go(t)}{\Go(k)}  \right) \right]\notag \\
    & - \frac{  \hat \pi_A (\pi_{\Gamma,0} - \hat \pi_{\Gamma}) +  (\pi_{A,0} - \hat \pi_A) (1-\hat \pi_{\Gamma}) \hat \pi_{\Gamma} + (\pi_{A,0} - \hat \pi_A) (\pi_{\Gamma,0} - \hat \pi_{\Gamma}) (1 - \hat \pi_{\Gamma})  }{\hat \pi_{\Gamma} \hat \pi_{A} (1-\pi_{\Gamma,0})} \bigg)  \bigg] \notag \\
\end{align}
Note that

\begin{align}\label{eq:trans_R2_2}
    \Eo \left[ \varphi_{\theta_T}(O, \theta_{T, \Po};\hat \eta, t, a) - \varphi_{\theta_T}(O;\hat \eta, t, a) \right] & = \hat \theta_{T,\hat \P} (t,a) - \theta_{T, \Po} (t,a)  \notag \\
    & + \left( \frac{\p_0}{\hat \p} - 1 \right) (\hat \theta_{T,\hat \P} (t,a) - \theta_{T, \Po} (t,a) )
\end{align}
Where we define the remainder term of the von-Mises expansion for $\theta_T(t)$, note that this term is the sum of products of difference between nuisances' true values and their estimates.

\[
\R_2(\hat \eta, \eta_0; \theta_T(t)) = (\ref{eq:trans_R2_1}) + (\ref{eq:trans_R2_2})
\]
Where all elements of the sum are products of the difference of true nuisances $\eta_0$ and their estimates $\hat \eta$. Finally we can show that the candidate influence function $\varphi_{\theta_T}(O;\eta_0, t)$ is the efficient influence function at every time point $t := 1,\dots,K$ by taking the remainder's pathwise derivative at $\epsilon = 0$. Note that we can simply flip the order of the parameters in the von-Mises expansion to simplify this differentiation

\begin{align*}
    \ppe \R_2(\eta_\epsilon, \eta_0; \theta_T(t)) \ate & = \ppe \R_2(\eta_0,\eta_\epsilon; \theta_T(t)) \ate \\
    & = \ppe  \Ee \hsum \bigg[ \frac{(1-\pi_{\Gamma,\epsilon})}{\p_0} \Se(m) ( h_0(m) - h_{\epsilon}(m)) \frac{\So(t)}{\So(m)} \notag \\
    & \times  \left( \frac{\pi_{A,\epsilon} \pi_{\Gamma,\epsilon} (1-\pi_{\Gamma,0}) \Ge(m) }{ \pi_{\Gamma,0} \pi_{A,0} (1-\pi_{\Gamma,\epsilon}) \Go(m)} -1 \right) \bigg]  + \left( \frac{\p_0}{\hat \p} - 1 \right) (\hat \theta_{T,\Po} (t,a) - \theta_{T, \Pe} (t,a) ) \ate \\
    & = \ppe  \sum_w \hsum \bigg[ \frac{(1-\pi_{\Gamma,\epsilon})}{\p_0} \Se(m) ( h_0(m) - h_{\epsilon}(m)) \frac{\So(t)}{\So(m)} \notag \\
    & \times  \left( \frac{\pi_{A,\epsilon} \pi_{\Gamma,\epsilon} (1-\pi_{\Gamma,0}) \Ge(m) }{ \pi_{\Gamma,0} \pi_{A,0} (1-\pi_{\Gamma,\epsilon}) \Go(m)} -1 \right) \bigg]  + \left( \frac{\p_0}{\hat \p} - 1 \right) (\hat \theta_{T,\Po} (t,a) - \theta_{T, \Pe} (t,a) ) \ate \\
    & = \sum_w \hsum \ppe \bigg[ \frac{(1-\pi_{\Gamma,\epsilon})}{\p_0} \Se(m) ( h_0(m) - h_{\epsilon}(m)) \frac{\So(t)}{\So(m)} \notag \\
    & \times  \left( \frac{\pi_{A,\epsilon} \pi_{\Gamma,\epsilon} (1-\pi_{\Gamma,0}) \Ge(m) }{ \pi_{\Gamma,0} \pi_{A,0} (1-\pi_{\Gamma,\epsilon}) \Go(m)} -1 \right) \bigg]  + \left( \frac{\p_0}{\hat \p} - 1 \right) (\hat \theta_{T,\Po} (t,a) - \theta_{T, \Pe} (t,a) ) \ate
\end{align*}
Under regularity conditions, enabling flipping differentiation and integration. Now simply expanding the expression and applying differentiation rules lead to a set of compensating terms, yields

\[
    \ppe \R_2(\eta_\epsilon, \eta_0; \theta_T(t)) \ate = \ppe \R_2(\eta_0,\eta_\epsilon; \theta_T(t)) \ate = 0
\]
This expands naturally to the difference contrast of treatment arms by differentiation rules. Thus $\varphi_{\theta_T}(O;\eta_0, t)$ is the efficient influence function of $\theta_{T,\Po}$ in the nonparametric model.


\subsubsection{Efficient influence function for generalization estimator}\label{sec:supp_gen_eif}

Let's start by taking the derivate $\theta_{G,\Pe}(t,a)$ at $\epsilon = 0$,

\begin{align}\label{eq:eif_gen_head}
    \ppe \theta_{G,\Pe}(t,a) \ate & = \ppe \E_0[ \Se(t \mid A = a,w,\gamma) ] \ate \notag \\
    & = \ppe \sum_w \left[ \Se(t \mid A = a,w,\gamma) p_{\epsilon}^* (w) \right] \ate \notag \\
    & = \sum_w \left[ \Se(t \mid A = a,w,\gamma) \ppe p_{\epsilon}^* (w) \ate + p_{\epsilon}^* (w) \ppe \Se(t \mid A = a,w,\gamma) \ate \right] \notag \\
    & = \sum_w \left[ \So(t \mid A = a,w,\gamma) \ppe p_{\epsilon}^* (w) \ate + p_0 (w) \ppe \Se(t \mid A = a,w,\gamma) \ate \right]
\end{align}
Where $\ppe p_{\epsilon}^* (w) \ate = \1 (W = w) - p_0(w)$, thus:

\begin{align*}
    \sum_w \left[ \So(t \mid A = a,w,\gamma) \ppe p_{\epsilon}^* (w) \ate \right] & = \So(t \mid A = a,w,\gamma) - \Eo \left[ \So(t \mid A = a,w,\gamma) \right] \\
    & = S_0(t \mid A = a,w,\gamma) - \theta_{G,a, \Po}(t)
\end{align*}
and
\begin{align*}
    \ppe \theta_{G,\Pe}(t,a) \ate & = \sum_w \left[ p_0 (w) \ppe \Se(t \mid A = a,w,\gamma) \ate \right] + \So(t \mid A = a,w,\gamma) - \theta_{G,\Po}(t,a)
\end{align*}
Where we already solved $\ppe \Se(t \mid A = a,w,\gamma) \ate$ in equations (\ref{eq:eif_surv_1}) and (\ref{eq:eif_surv_2}), which we can plug-into equation (\ref{eq:eif_gen_head}) and summing over all $w$'s yields the following treatment specific derivative yielding:

\begin{align*}
    \ppe \theta_{G,\Pe}(t,a) \ate & = \frac{\1(A=a,\Gamma =\gamma) }{\Po(A =a \mid W,\gamma) \Po(\Gamma =1 \mid W)} \So(t \mid a,w,\gamma) \\
    & \times \hsum \left( \frac{\1(I_m = 1)[ h_0(m \mid a,w,\gamma) - \1(L_m = 1)]}{\So(m\mid a,w,\gamma) \Go(m \mid a,w,\gamma)} \right) \\
    & + \1(\Gamma =0) \left( \So(t \mid a,w,\gamma) - \theta_{T,\Po}(t,a)  \right) 
\end{align*}
Finally by combining both treatment branches and some observed $A = a$ we get the candidate influence function:

\begin{align}\label{eq:cand_gen_eif}
    \varphi_{\theta_G}(O;\eta_0, t) = & \ppe \theta_{G,\Pe}(t) \ate = \ppe \theta_{G,\Pe}(t,A=1) \ate - \ppe \theta_{G,\Pe}(t,A=0) \ate \notag \\
    &= \frac{\1(\Gamma = 1) }{\pi_{\Gamma,0} }  \left( \frac{\1(A = 1)}{\pi_{A,0}} - \frac{\1 (A = 0)}{1 -\pi_{A,0}} \right) \D(t) +\left( S_0^*(t) - \theta_{G, \Po}(t) \right) 
\end{align}
The next step is to find the remainder of this candidate influence function in the von-Mises expansion. Note that, once again we can focus on a single treatment arm $A=a$ before extending to the difference of the treatment arms, and we can make heavy use of the results of the previous section \ref{sec:supp_trans_eif}. As this case is simply a modified case, with equations (\ref{eq:cross_R}) and (\ref{eq:trans_R2_1}) along with applying Lemma \ref{supp_lemma:seq} in similar ways we get 

\begin{align}\label{eq:gen_R2}
    \R_2(\hat \eta, \eta_0; \theta_G(t)) & = \Eo \bigg[ \hsum \So(m) (\hat h(m) - h_0(m)) \frac{\hat \S(t)}{\hat \S(m)} \notag \\
    & \times  \bigg( \frac{\pi_{A,0} \pi_{\Gamma,0} }{\hat \pi_{\Gamma} \hat \pi_{A} \hat \G(m)} \left[ \sum_{k = 0}^{m -1} \left( G(k) (\hat g_{C}(k) - g_{C,0}(k)) \frac{\Go(t)}{\Go(k)}  \right) \right]\notag \\
    & - \frac{ (\pi_{A,0} - \hat \pi_A) (\pi_{\Gamma, 0} - \hat \pi_{\Gamma}) + \hat \pi_A (\pi_{\Gamma, 0} - \hat \pi_{\Gamma}) + \hat \pi_{\Gamma} (\pi_{A,0} - \hat \pi_A)  }{\hat \pi_{\Gamma} \hat \pi_{A} } \bigg)  \bigg] \notag \\
\end{align}
Where all elements of the sum are products of the difference of true nuisances $\eta_0$ and their
estimates $\hat \eta$. Finally we can show that the candidate influence function $\varphi_{\theta_T}
(O;\eta_0, t)$ is the efficient influence function at every time point $t := 1,\dots,K$ by taking
the remainder's pathwise derivative at $\epsilon = 0$. Note that we can simply flip the order of
the parameters in the von-Mises expansion to simplify this differentiation and using a simplified version of the remainder term:

\begin{align*}
    \ppe \R_2(\hat \eta, \eta_0; \theta_G(t)) \ate & = \ppe \R_2(\eta_0, \hat \eta; \theta_G(t)) \ate  \\
    & = \ppe \Ee \left[ \hsum \frac{\So(t)}{\So(m-1)} \Se(m-1) (\he(m) - h_0(m)) \left( \frac{\pi_{A,\epsilon} \pi_{\Gamma,\epsilon}}{\pi_{A,0} \pi_{\Gamma,0}} \frac{\Ge(t)}{\G_0(t)} - 1 \right) \right] \ate \\
    & = \hsum \sum_W \ppe \left[ \frac{\So(t)}{\So(m-1)} \Se(m-1) (\he(m) - h_0(m)) \left( \frac{\pi_{A,\epsilon} \pi_{\Gamma,\epsilon}}{\pi_{A,0} \pi_{\Gamma,0}} \frac{\Ge(t)}{\G_0(t)} - 1 \right) \right]  \ate
\end{align*}
Expanding using product rule yields the following expression with compensating terms:

\begin{align*}
    \ppe \R_2(\hat \eta, \eta_0; \theta_G(t)) \ate & = \ppe \R_2(\eta_0, \hat \eta; \theta_G(t)) \ate = h_0(t) \ppe \Se (t-1) \ate - \So (t-1) \ppe \he(t) \ate \\
    & - h_0(t) \ppe \Se (t-1) \ate + h_0(t) \ppe \Se(t-1) \ate + \So(t-1) \ppe \he(t) \ate \\
    & + \frac{\So(t-1) h_0(t)}{\pi_{A,0}} \ppe \pi_{A,\epsilon} \ate + \frac{\So(t-1) h_0(t)}{\Go (t)} \ppe \Ge (t) \ate + \frac{\So(t-1) h_0(t)}{\pi_{\Gamma,0}} \ppe \pi_{\Gamma,\epsilon} \ate \\
    & - h_0(t) \ppe \Se(t-1) \ate - \frac{\So(t-1) h_0(t)}{g_{A,0}} \ppe g_{A,\epsilon} \ate \\
    & - \frac{\So(t-1) h_0(t)}{\Go(t)} \ppe \Ge(t) \ate - \frac{\So(t-1) h_0(t)}{\pi_{\Gamma,0}} \ppe \pi_{\Gamma,\epsilon} \ate = 0 
\end{align*}
This expands naturally to the difference contrast of treatment arms by differentiation rules. Thus $\varphi_{\theta_G}(O;\eta_0, t)$ is the efficient influence function of $\theta_{G,\Po}$ in the nonparametric model.


\subsubsection{Efficient influence function for transport estimator leveraging the effect modifier subset}\label{sec:supp_transeff_eif}

Again, we can rewrite our parameter as a difference of the treatment arm of the parameter, let $\gamma = 1$ unless specified otherwise

\begin{align*}
    \lambda_{T,\Po} (t) &= \Eo \left[ \Eo \left[ \f_0(t,V) \mid  Z \right] \mid \Gamma = 0 \right] \\
    & = \Eo \left[ \Eo \left[ \So(t \mid A = 1, w, \gamma) - \So(t \mid A = 0, w, \gamma) \mid Z \right] \mid \Gamma = 0  \right] \\
    & = \Eo \left[ \Eo \left[ \So(t \mid A = 1, w, \gamma)\mid Z \right] \mid \Gamma = 0  \right] - \Eo \left[ \Eo \left[ \So(t \mid A = 0, w, \gamma) \mid Z \right] \mid \Gamma = 0  \right] \\
    & = \lambda_{T,\Po} (t,A=1) - \lambda_{T,\Po} (t,A=0)
\end{align*}
Where

\begin{align*}
    \ppe \lambda_{T,\Pe} (t,a) \ate & = \ppe \Eo \left[ \Eo \left[ \Se (t \mid a, w,\gamma) \mid Z \right] \mid \Gamma =0 \right] \ate \\
    & = \ppe \sum_{w_0} \left[ \sum_{w_1}  \Pe(w_1 \mid Z) \Se(t\mid a ,w,\gamma)  \right] \Pe (w_0 \mid \Gamma =0) \ate
\end{align*}
with

\[ 
\ppe \Pe(w_1 \mid Z) \ate = \frac{\1 (Z = z)}{ \Po (Z =z )} (\1(W = w_1) - \Po (w_1 \mid Z = z))
\]
\[
\ppe \Pe(w_0 \mid \Gamma =0) \ate = \frac{\1 (\Gamma = 0)}{ \Po (\Gamma = 0)} (\1(W = w_0) - \Po (w_0 \mid \Gamma = 0))
\]
and $\ppe \Se(t\mid a ,w,\gamma) \ate$ having been solved in equation (\ref{eq:eif_surv_1}) and (\ref{eq:eif_surv_2}), thus we have for a single treatment arm

\begin{align*}
    \ppe \lambda_{T,\Pe} (t,a) \ate & = \frac{1}{\Po (\Gamma = 0)} \bigg[ \hsum \frac{\1(\Gamma =1, A = a) \Po(\Gamma = 0\mid Z)}{\Po(A = a \mid W, \Gamma = 1) \Po(\Gamma =1 \mid Z)} \\
    & \times \So(t) \frac{\1(I_m = 1) (h_0(m) - \1(\L_m = 1))}{\So(m ) \Go(m)}  \\
    & + \Po (\Gamma =0 \mid Z) \left[ \So(t) - \Eo\left[ \So(t) \mid Z \right]  \right]  + \1 (\Gamma = 0) \left[ \Eo\left[ \So(t) \mid Z \right] - \lambda_{T,\Po} (t,a)  \right] \bigg]
\end{align*}
Now combining both treatment arms under Assumptions~\ref{ass:param} through~\ref{ass:hem} yields

\begin{align*}
    \varphi_{\lambda_T}(O;\eta^*_0, t) & = \ppe \lambda_{T,\Pe} (t) \ate = \ppe \lambda_{T,\Pe} (t,A=1) \ate - \ppe \lambda_{T,\Pe} (t,A=0) \ate \notag \\
    & = \frac{1}{\p_0} \bigg[ \frac{\1(\Gamma = 1) (1-\pi^*_{\Gamma,0}) }{\pi^*_{\Gamma,0} }  \left( \frac{\1(A = 1)}{\pi_{A,0}} - \frac{\1 (A = 0)}{1 -\pi_{A,0}} \right) \D(t) \notag \\ 
    & + (1-\pi^*_{\Gamma,0}) \left( \fo(t) - \Eo \left[ \fo(t) \mid Z \right] \right) + \1(\Gamma = 0) \left( \Eo \left[ \fo(t) \mid Z \right] - \lambda_{T, \Po}(t) \right) \bigg]
\end{align*}
The next step is to find the remainder of this candidate influence function in the von-Mises expansion. Let us first consider the cross-term where nuisance parameters are estimated under $\hat \P$ but the true parameter $\lambda_{T,\Po}(t) $ remains the final term of the equation above. The portions highlighted in \textcolor{blue}{blue} below are the terms that are written as products of difference between true and estimated nuisances:

\begin{align*}
    &\Eo \left[ \varphi_{\lambda_T}(O, \lambda_{T, \Po};\hat \eta^*, t) \right] = \\
    &\Eo \bigg[ \frac{1}{\hat \p} \bigg( \hsum \frac{\1(\Gamma=1) (1 - \hat \pi_{\Gamma}^*) }{\hat \pi_{\Gamma}^* }  \left( \frac{\1(A=1)}{\hat \pi_A} - \frac{\1(A=0)}{\hat \pi_A}  \right) \frac{\1(I_m =1) \hat \S(t) (\hat h(m) - \1(L_m = 1))}{\hat \S (m) \hat \G(m)}   \\
    & + (1 - \hat \pi_{\Gamma}^*) (\hat \f(t) - \hat \E [ \hat \f(t) \mid Z]) + \1 (\Gamma = 0) ( \hat \E [\hat \f (t) \mid Z] - \lambda_{T,\Po}(t) ) \bigg)   \bigg]  \\
    & = \Eo \bigg[ \frac{1}{\hat \p} \bigg( \hsum \frac{\pi_{\Gamma,0}^* (1 - \hat \pi_{\Gamma}^*) }{\hat \pi_{\Gamma}^* }  \left( \frac{\pi_{A,0}}{\hat \pi_A} - \frac{1-\pi_{A,0}}{\hat \pi_A}  \right) \frac{\hat \S(t) \So(m) \Go(m) (\hat h(m) - h_0(m))}{\hat \S (m) \hat \G(m)}  \\
    & + (1 - \hat \pi_{\Gamma}^*) (\hat \f(t) - \hat \E [ \hat \f(t) \mid Z]) + (1-\pi_{\Gamma,0}^*) ( \hat \E [\hat \f (t) \mid Z] - \lambda_{T,\Po}(t))  \bigg)    \bigg]  \\
    & = \Eo \left[ \frac{1}{\hat \p}  \hsum \frac{\pi_{\Gamma,0}^* (1 - \hat \pi_{\Gamma}^*) }{\hat \pi_{\Gamma}^* }  \left( \frac{\pi_{A,0}}{\hat \pi_A} - \frac{1-\pi_{A,0}}{\hat \pi_A}  \right) \frac{\hat \S(t) \So(m)  (\hat h(m) - h_0(m))}{\hat \S (m) } \left( \frac{G(m) - \hat G(m) + \hat G(m)}{\hat G(m)} \right) \right]  \\
    & + \Eo \left[ \frac{1 - \hat \pi_{\Gamma}^*}{\hat \p} (\hat \f (t) - \f_0(t)) \right]   \\
    & + \textcolor{blue}{\Eo \left[ \frac{\pi_{\Gamma,0} - \hat \pi_{\Gamma}}{\hat \p} \left( \Eo [ \hat \f \mid Z] - \hat \E [ \hat \f \mid Z] \right)  \right]} \tag{i}\label{eq:R2_transeff_i} \\
    & = \Eo \left[ \frac{1}{\hat \p}  \hsum \frac{\pi_{\Gamma,0}^* (1 - \hat \pi_{\Gamma}^*) }{\hat \pi_{\Gamma}^* }  \left( \frac{\pi_{A,0}}{\hat \pi_A} - \frac{1-\pi_{A,0}}{\hat \pi_A}  \right) \frac{\hat \S(t) \So(m)  (\hat h(m) - h_0(m))}{\hat \S (m) } \left( \sum_{k = 0}^{m-1} \frac{\Go(k)}{\hat \G(k)} (g_{C,0}(k) - \hat g_C(k)) + 1 \right) \right]  \\
    & + \Eo \left[ \frac{1 - \hat \pi_{\Gamma}^*}{\hat \p} (\hat \f (t) - \f_0(t)) \right] + (\ref{eq:R2_transeff_i}) \\
    & = \textcolor{blue}{\Eo \left[ \frac{1}{\hat \p}  \hsum \frac{\pi_{\Gamma,0}^* (1 - \hat \pi_{\Gamma}^*) }{\hat \pi_{\Gamma}^* }  \left( \frac{\pi_{A,0}}{\hat \pi_A} - \frac{1-\pi_{A,0}}{\hat \pi_A}  \right) \frac{\hat \S(t) \So(m)  (\hat h(m) - h_0(m))}{\hat \S (m) } \left( \sum_{k = 0}^{m-1} \frac{\Go(k)}{\hat \G(k)} (g_{C,0}(k) - \hat g_C(k)) \right) \right]}  \tag{ii}\label{eq:R2_transeff_ii} \\
    & + \Eo \left[ \frac{1}{\hat \p}  \hsum \frac{\pi_{\Gamma,0}^* (1 - \hat \pi_{\Gamma}^*) }{\hat \pi_{\Gamma}^* }  \left( \frac{\pi_{A,0}}{\hat \pi_A} - \frac{1-\pi_{A,0}}{\hat \pi_A}  \right) \frac{\hat \S(t) \So(m)  (\hat h(m) - h_0(m))}{\hat \S (m) }  \right] + \Eo \left[ \frac{1 - \hat \pi_{\Gamma}^*}{\hat \p} (\hat \f (t) - \f_0(t)) \right] + (\ref{eq:R2_transeff_i}) \\
    & = \Eo \left[ \frac{1}{\hat \p}  \frac{\pi_{\Gamma,0}^* (1 - \hat \pi_{\Gamma}^*) }{\hat \pi_{\Gamma}^* }  \left( \frac{\pi_{A,0}}{\hat \pi_A} - \frac{1-\pi_{A,0}}{\hat \pi_A}  \right) \left( \hprod (1 - h_0(m)) - \hprod (1 - \hat h(m)) \right)  \right] \\
    & + \Eo \left[ \frac{1 - \hat \pi_{\Gamma}^*}{\hat \p} (\hat \f (t) - \f_0(t)) \right] + (\ref{eq:R2_transeff_i}) + (\ref{eq:R2_transeff_ii}) \\
    & = \Eo \left[ \frac{1}{\hat \p}  \frac{\pi_{\Gamma,0}^* (1 - \hat \pi_{\Gamma}^*) }{\hat \pi_{\Gamma}^* }  \left( \frac{\pi_{A,0}}{\hat \pi_A} - \frac{1-\pi_{A,0}}{\hat \pi_A}  \right) \left( A \f(t) + \g(t) - A \hat \f - \hat \g (t) \right)  \right] \\
    & + \Eo \left[ \frac{1 - \hat \pi_{\Gamma}^*}{\hat \p} (\hat \f (t) - \f_0(t)) \right] + (\ref{eq:R2_transeff_i}) + (\ref{eq:R2_transeff_ii}) 
\end{align*} 

\begin{align*}
    & = \Eo \left[ \frac{\pi_{\Gamma,0}^* (1 - \hat \pi_{\Gamma}^*) \pi_{A,0} }{\hat \p \hat \pi_{\Gamma}^* \hat \pi_A} \left( \f_0(t) - \hat \f(t) \right)    \right] + \textcolor{blue}{ \Eo \left[ \frac{\pi_{\Gamma,0}^* (1 - \hat \pi_{\Gamma}^*) }{\hat \p \hat \pi_{\Gamma}^* }  \left( \frac{\pi_{A,0}}{\hat \pi_A} - \frac{1-\pi_{A,0}}{\hat \pi_A}  \right) \left( \g(t) - \hat \g (t) \right)  \right] } \tag{iii}\label{eq:R2_transeff_iii} \\
    & + \Eo \left[ \frac{1 - \hat \pi_{\Gamma}^*}{\hat \p} (\hat \f (t) - \f_0(t)) \right] + (\ref{eq:R2_transeff_i}) + (\ref{eq:R2_transeff_ii}) \\
    & = \Eo \left[ \left( \frac{\pi_{\Gamma,0}^* (1 - \hat \pi_{\Gamma}^*) \pi_{A,0} }{\hat \p \hat \pi_{\Gamma}^* \hat \pi_A} - \frac{1 - \hat \pi_{\Gamma}^*}{\hat \p}  \right) \left( \f_0(t) - \hat \f(t) \right) \right] + (\ref{eq:R2_transeff_i}) + (\ref{eq:R2_transeff_ii}) + (\ref{eq:R2_transeff_iii}) \\
    & = \textcolor{blue}{ \Eo \left[ \left( \frac{\pi_{\Gamma,0}^* (1 - \hat \pi_{\Gamma}^*)  }{\hat \p \hat \pi_{\Gamma}^*} \left(\frac{\pi_{A,0}}{\hat \pi_A} -1 \right) \right) \left( \f_0(t) - \hat \f(t) \right) \right] } \tag{iv}\label{eq:R2_transeff_iv} \\
    & + \textcolor{blue}{\Eo \left[ \frac{\pi_{\Gamma, 0}^* - \hat \pi_{\Gamma}^*}{\hat \pi_{\Gamma}^*} \left( \f_0(t) - \hat \f(t) \right)\right] } \tag{v}\label{eq:R2_transeff_v} + (\ref{eq:R2_transeff_i}) + (\ref{eq:R2_transeff_ii}) + (\ref{eq:R2_transeff_iii})
\end{align*}
Note that

\begin{align*}
    \Eo \left[ \varphi_{\lambda_T}(O, \lambda_{T, \Po};\hat \eta^*, t) - \varphi_{\lambda_T}(O;\hat \eta^*, t) \right] & = \hat \lambda_{T,\hat \P} (t) - \lambda_{T, \Po} (t)  \notag \\
    & + \textcolor{blue}{ \left( \frac{\p_0}{\hat \p} - 1 \right) (\hat \lambda_{T,\hat \P} (t,a) - \lambda_{T, \Po} (t,a) )} \tag{vi}\label{eq:R2_transeff_vi}
\end{align*}
Where we define the remainder term of the von-Mises expansion for $\lambda_T(t)$, note that this term is the sum of products of difference between nuisances' true values and their estimates.

\[
\R_2(\hat \eta^*, \eta_0^*; \lambda_T(t)) = (\ref{eq:R2_transeff_i}) + (\ref{eq:R2_transeff_ii}) + (\ref{eq:R2_transeff_iii}) + (\ref{eq:R2_transeff_iv}) + (\ref{eq:R2_transeff_v}) + (\ref{eq:R2_transeff_vi}) 
\]
Where all elements of the sum are products of the difference of true nuisances $\eta^*_0$ and their estimates $\hat \eta^*$. Finally we can show that the candidate influence function $\varphi_{\lambda_T}(O;\eta_0, t)$ is the efficient influence function at every time point $t := 1,\dots,K$ by taking the remainder's pathwise derivative at $\epsilon = 0$. Note that we can simply flip the order of the parameters in the von-Mises expansion to simplify this differentiation, then under regularity conditions we can flip the order of differentiation and integrations, expand each of the elements of the sum and apply differentiation rules to obtain the desired result. 

\[
    \ppe \R_2(\eta_\epsilon^*, \eta_0^*; \lambda_T(t)) \ate = \ppe \R_2(\eta_0^*,\eta_\epsilon^*; \lambda_T(t)) \ate = 0
\]
This expands naturally to the difference contrast of treatment arms by differentiation rules. Thus $\varphi_{\lambda_T}(O;\eta_0^*, t)$ is the efficient influence function of $\lambda_{T,\Po}$ in the nonparametric model.


\subsubsection{Efficient influence function for generalizaton estimator leveraging the effect modifier subset}\label{sec:supp_geneff_eif}

We write the parameter using the tower rule as

\begin{align*}
    \lambda_{G,\Po} (t) &= \Eo \left[ \f_0(t,V) \right] = \Eo \left[ \Eo \left[ \f_0(t,V) \mid Z \right] \right] \\
    & = \Eo \left[ \So(t \mid A = 1, w, \gamma) - \So(t \mid A = 0, w, \gamma) \right] \\
    & = \lambda_{G,\Po} (t,A=1) - \lambda_{G,\Po} (t,A=0)
\end{align*}
where the second equality uses only the tower property of conditional expectation and does not require Assumption~\ref{ass:hem}. For a single treatment arm, $\lambda_{G,\Po}(t,a) = \sum_z \Eo[\So(t \mid a, w, \gamma) \mid Z = z] \Po(z)$. Taking the Gateaux derivative through this representation:

\begin{align*}
    \ppe \lambda_{G,\Pe} (t,a) \ate & = \ppe \sum_z \left[ \sum_{w,\gamma}  \Se(t\mid a ,w,\gamma) \Pe(w,\gamma \mid Z = z)  \right] \Pe (z) \ate
\end{align*}
with

\[
\ppe \Pe(w,\gamma \mid Z) \ate = \frac{\1 (Z = z)}{ \Po (Z =z )} (\1(W = w, \Gamma = \gamma) - \Po (w, \gamma \mid Z = z))
\]
\[
\ppe \Pe(z) \ate = \1(Z = z) - \Po(z)
\]
and $\ppe \Se(t\mid a ,w,\gamma) \ate$ having been solved in equation (\ref{eq:eif_surv_1}) and (\ref{eq:eif_surv_2}). After applying the product rule to $\Pe(w,\gamma|Z)$, $\Se(t|a,w,\gamma)$, and $\Pe(z)$, and summing over all $w$, $\gamma$, and $z$, we obtain for a single treatment arm:

\begin{align*}
    \ppe \lambda_{G,\Pe} (t,a) \ate & = \hsum \frac{\1(\Gamma = \gamma, A = a)}{\Po(A = a \mid W, \Gamma = \gamma) \Po(\Gamma =\gamma \mid Z)} \So(t) \frac{\1(I_m = 1) (h_0(m) - \1(\L_m = 1))}{\So(m ) \Go(m)}  \\
    & + \left[ \So(t \mid a, w, \gamma) - \Eo\left[ \So(t \mid a) \mid Z \right]  \right]  + \left[ \Eo\left[ \So(t \mid a) \mid Z \right] - \lambda_{G,\Po} (t,a)  \right]
\end{align*}
The last two terms combine to give $\So(t \mid a, w, \gamma) - \lambda_{G,\Po}(t,a)$. Combining both treatment arms under Assumptions~\ref{ass:param} through~\ref{ass:tem} yields

\begin{align*}
    \varphi_{\lambda_G}(O;\eta^*_0, t) & = \ppe \lambda_{G,\Pe} (t) \ate = \ppe \lambda_{G,\Pe} (t,A=1) \ate - \ppe \lambda_{G,\Pe} (t,A=0) \ate \notag \\
    & = \frac{\1(\Gamma = 1) }{\pi^*_{\Gamma,0} }  \left( \frac{\1(A = 1)}{\pi_{A,0}} - \frac{\1 (A = 0)}{1 -\pi_{A,0}} \right) \D(t) + \left( \fo(t) - \lambda_{G, \Po}(t) \right)
\end{align*}
which matches equation (\ref{eq:gen_eff_eif}). Note that $\pi^*_{\Gamma,0} = \Po(\Gamma = \gamma \mid Z)$ enters the IPW term through the conditioning on $Z$ in the representation $\lambda_G = \Eo[\Eo[\fo \mid Z]]$, even though the estimand itself equals the marginal $\Eo[\fo(t,V)]$. This is the source of the efficiency gain relative to $\varphi_{\theta_G}$, which uses $\pi_{\Gamma,0}(\Gamma \mid W)$.

The next step is to find the remainder of this candidate influence function in the von-Mises expansion. The structure is analogous to the base generalization estimator $\theta_G$ (Section \ref{sec:supp_gen_eif}) but with $\pi^*_{\Gamma,0}(\Gamma \mid Z)$ replacing $\pi_{\Gamma,0}(\Gamma \mid W)$. Let us consider the cross-term where nuisance parameters are estimated under $\hat \P$ but the true parameter $\lambda_{G,\Po}(t)$ remains:

\[
\varphi_{\lambda_G}(O, \lambda_{G,\Po};\hat \eta^*, t, a) = \frac{\1(\Gamma = 1, A=a)}{\hat \pi^*_\Gamma \hat \pi_A} \hat \D(t) + \hat \S(t \mid a) - \lambda_{G,\Po}(t,a)
\]

Using iterated expectations and the same algebraic steps as in equation (\ref{eq:gen_R2}) for $\theta_G$, replacing $\pi_{\Gamma,0}$ by $\pi^*_{\Gamma,0}$ throughout:

\begin{align*}
    \R_2(\hat \eta^*, \eta_0^*; \lambda_G(t)) & = \Eo \bigg[ \hsum \So(m) (\hat h(m) - h_0(m)) \frac{\hat \S(t)}{\hat \S(m)} \\
    & \times  \bigg( \frac{\pi_{A,0} \pi^*_{\Gamma,0}}{\hat \pi^*_{\Gamma} \hat \pi_{A} \hat \G(m)} \left[ \sum_{k = 0}^{m -1} \left( G(k) (\hat g_{C}(k) - g_{C,0}(k)) \frac{\Go(t)}{\Go(k)}  \right) \right]\\
    & + \frac{ (\pi_{A,0} - \hat \pi_A) (\pi^*_{\Gamma, 0} - \hat \pi^*_{\Gamma}) + \hat \pi_A (\pi^*_{\Gamma, 0} - \hat \pi^*_{\Gamma}) + \hat \pi^*_{\Gamma} (\pi_{A,0} - \hat \pi_A)  }{\hat \pi^*_{\Gamma} \hat \pi_{A} } \bigg)  \bigg]
\end{align*}
All terms are products of $(\hat h(m) - h_0(m))$ with differences in $(g_C, \pi_A, \pi^*_\Gamma)$. Flipping the order of the parameters and differentiating at $\epsilon = 0$ under regularity conditions yields

\[
    \ppe \R_2(\eta_\epsilon^*, \eta_0^*; \lambda_G(t)) \ate = \ppe \R_2(\eta_0^*,\eta_\epsilon^*; \lambda_G(t)) \ate = 0
\]
This extends to the treatment contrast by differentiation rules. Thus $\varphi_{\lambda_G}(O;\eta_0^*, t)$ is the efficient influence function of $\lambda_{G,\Po}$ in the nonparametric model.


\subsection{Proof of Theorem \ref{thm:vonmises}}

For $\theta_T(\tau)$:

Let us first note that $\left( \frac{\p_0}{\hat \p} - 1 \right) (\hat \theta_{T,\hat \P} (t,a) - \theta_{T, \Po} (t,a) ) = o_P(n^{-1/2})$ because $\hat \p$ is an empirical average, thus $\frac{\p_0}{\hat\p}-1 = O_P(n^{-1/2})$ and $\hat \theta_{T,\hat \P} (t,a) \underset{P}{\to} \theta_{T, \Po} (t,a)$ under consistency of the nuisance parameters. And will be ignored in the proof below:

\begin{align*}
    \R_2(\hat \eta, \eta_0; \theta_T(\tau)) & = \Eo \bigg[ \frac{(1-\pi_{\Gamma,0})}{\hat \p} \hsumtau \So(m) (\hat h(m) - h_0(m)) \frac{\hat \S(t)}{\hat \S(m)} \\
    & \times  \bigg( \frac{\pi_{A,0} \pi_{\Gamma,0} (1-\hat \pi_{\Gamma}) }{\hat \pi_{\Gamma} \hat \pi_{A} (1-\pi_{\Gamma,0}) \hat \G(m)} \left[ \sum_{k = 0}^{m -1} \left( \hat \G(k) (\hat g_{C}(k) - g_{C,0}(k)) \frac{\Go(t)}{\Go(k)}  \right) \right] \\
    & - \frac{  \hat \pi_A (\pi_{\Gamma,0} - \hat \pi_{\Gamma}) +  (\pi_{A,0} - \hat \pi_A) (1-\hat \pi_{\Gamma}) \hat \pi_{\Gamma} + (\pi_{A,0} - \hat \pi_A) (\pi_{\Gamma,0} - \hat \pi_{\Gamma}) (1 - \hat \pi_{\Gamma})  }{\hat \pi_{\Gamma} \hat \pi_{A} (1-\pi_{\Gamma,0})} \bigg)  \bigg] \\
    & = \hsumtau \sum_{k = 0}^{m -1} \Eo \left[ \frac{\So(m) \hat \S(t)}{\hat \S(m)} \frac{\pi_{A,0} \pi_{\Gamma,0} (1-\hat \pi_{\Gamma}) \hat \G(k) \Go(t)}{\hat \pi_{\Gamma} \hat \pi_{A} (1-\pi_{\Gamma,0}) \hat \G(m) \Go(k)}  (\hat h(m) - h_0(m)) (\hat g_{C}(k) - g_{C,0}(k))   \right] \\
    & + \hsumtau \sum_{k = 0}^{m -1} \Eo \left[ \frac{\So(m) \hat \S(t)}{\hat \S(m)}  \frac{\hat \pi_A ( \pi_{\Gamma,0} - \hat \pi_{\Gamma} ) + \pi_{\Gamma,0} ( 1- \hat \pi_{\Gamma}) }{\hat \pi_{\Gamma} \hat \pi_{A} (1-\pi_{\Gamma,0})}      (\hat h(m) - h_0(m)) (\pi_{A,0} - \hat \pi_A)  \right] \\
    & + \hsumtau \sum_{k = 0}^{m -1} \Eo \left[ \frac{\So(m) \hat \S(t)}{\hat \S(m)}  \frac{ (\hat \pi_{\Gamma} + \hat \pi_A) (1 - \hat \pi_{\Gamma}) (\pi_{A,0} - \hat \pi_A)   }{\hat \pi_{\Gamma} \hat \pi_{A} (1-\pi_{\Gamma,0})}   (\hat h(m) - h_0(m))  ( \pi_{\Gamma,0} - \hat \pi_{\Gamma} )  \right]
\end{align*}

Getting the supremum of all constant values and bounding them in each equation and then applying Cauchy-Schwarz inequality yields:

\begin{align*}
    \R_2(\hat \eta, \eta_0; \theta_T(\tau)) & \leq \hsumtau \sum_{k = 0}^{m -1} \alpha_1 \Eo \left[ (\hat h(m) - h_0(m)) \times  (\hat g_{C}(k) - g_{C,0}(k)) \right] \\
    & + \hsumtau \sum_{k = 0}^{m -1} \alpha_2 \Eo \left[ (\hat h(m) - h_0(m)) \times  (\hat \pi_A - \pi_{A,0}) \right] \\
    & + \hsumtau \sum_{k = 0}^{m -1} \alpha_3 \Eo \left[ (\hat h(m) - h_0(m)) \times  (\hat \pi_{\Gamma} - \pi_{\Gamma,0}) \right] \\
    & \leq \hsumtau \sum_{k = 0}^{m -1} \alpha_1 ||\hat h(m) - h_0(m)|| \times  ||\hat g_{C}(k) - g_{C,0}(k)|| \\
    & + \hsumtau \sum_{k = 0}^{m -1} \alpha_2 ||\hat h(m) - h_0(m)|| \times  ||\hat \pi_A - \pi_{A,0}|| \\
    & + \hsumtau \sum_{k = 0}^{m -1} \alpha_3 ||\hat h(m) - h_0(m)) \times  ||\hat \pi_{\Gamma} - \pi_{\Gamma,0}|| \\
    & = \hsumtau \sum_{k = 0}^{m -1} ||\hat h(m) - h_0(m)|| \left( \alpha_1 ||\hat g_{C}(k) - g_{C,0}(k)|| + \alpha_2 ||\hat \pi_A - \pi_{A,0}|| + \alpha_3 ||\hat \pi_{\Gamma} - \pi_{\Gamma,0}||  \right) \\
    & = \hsumtau \sum_{k = 0}^{m -1} O_{\hat \P} \left[ \|\hat h(m) - h_0(m)\| \left( \|\hat g_{C}(k) - g_{C,0}(k)\| + \|\hat \pi_A - \pi_{A,0}\| + \|\hat \pi_{\Gamma} - \pi_{\Gamma,0}\|  \right) \right]
\end{align*}
which matches equation (\ref{eq:R2_base}).

\textbf{For $\theta_G(\tau)$:} The remainder $\R_2(\hat \eta, \eta_0; \theta_G(\tau))$ was derived in equation (\ref{eq:gen_R2}) and takes the same form as $\theta_T$, without the prefactor $(1-\pi_{\Gamma,0})/\hat \p$ and with a simplified propensity structure. Applying Lemma \ref{supp_lemma:seq} to $\Go(m) - \hat \G(m)$ and grouping the numerator $\hat \pi_A (\pi_{\Gamma,0} - \hat \pi_\Gamma) + \hat \pi_\Gamma (\pi_{A,0} - \hat \pi_A) + (\pi_{A,0} - \hat \pi_A)(\pi_{\Gamma,0} - \hat \pi_\Gamma)$ into a term carrying $(\pi_{A,0} - \hat \pi_A)$ and a term carrying $(\pi_{\Gamma,0} - \hat \pi_\Gamma)$ yields, as for $\theta_T$,

\begin{align*}
    \R_2(\hat \eta, \eta_0; \theta_G(\tau)) & = \hsumtau \sum_{k = 0}^{m -1} \Eo \left[ \frac{\So(m) \hat \S(t)}{\hat \S(m)} \frac{\pi_{A,0} \pi_{\Gamma,0} \hat \G(k) \Go(t)}{\hat \pi_{\Gamma} \hat \pi_{A} \hat \G(m) \Go(k)}  (\hat h(m) - h_0(m)) (\hat g_{C}(k) - g_{C,0}(k))   \right] \\
    & + \hsumtau \sum_{k = 0}^{m -1} \Eo \left[ \frac{\So(m) \hat \S(t)}{\hat \S(m)}  \frac{1}{\hat \pi_{A}}      (\hat h(m) - h_0(m)) (\pi_{A,0} - \hat \pi_A)  \right] \\
    & + \hsumtau \sum_{k = 0}^{m -1} \Eo \left[ \frac{\So(m) \hat \S(t)}{\hat \S(m)}  \frac{ \pi_{A,0}   }{\hat \pi_{\Gamma} \hat \pi_{A} }   (\hat h(m) - h_0(m))  ( \pi_{\Gamma,0} - \hat \pi_{\Gamma} )  \right]
\end{align*}

Getting the supremum of all constant values and bounding them in each equation and then applying Cauchy-Schwarz inequality yields:

\begin{align*}
    \R_2(\hat \eta, \eta_0; \theta_G(\tau)) & \leq \hsumtau \sum_{k = 0}^{m -1} \|\hat h(m) - h_0(m)\| \left( \beta_1 \|\hat g_{C}(k) - g_{C,0}(k)\| + \beta_2 \|\hat \pi_A - \pi_{A,0}\| + \beta_3 \|\hat \pi_{\Gamma} - \pi_{\Gamma,0}\|  \right) \\
    & = \hsumtau \sum_{k = 0}^{m -1} O_{\hat \P} \left[ \|\hat h(m) - h_0(m)\| \left( \|\hat g_{C}(k) - g_{C,0}(k)\| + \|\hat \pi_A - \pi_{A,0}\| + \|\hat \pi_{\Gamma} - \pi_{\Gamma,0}\|  \right) \right]
\end{align*}
which matches equation (\ref{eq:R2_base}).

\textbf{For $\lambda_T(\tau)$:} The term (\ref{eq:R2_transeff_vi}), $\left( \frac{\p_0}{\hat \p} - 1 \right) (\hat \lambda_{T,\hat \P}(\tau) - \lambda_{T,\Po}(\tau)) = o_\P(n^{-1/2})$ by the same argument as for $\theta_T$, and is ignored below. Collecting the remaining terms (\ref{eq:R2_transeff_i})--(\ref{eq:R2_transeff_v}),

\[
\R_2(\hat \eta^*, \eta_0^*; \lambda_T(\tau)) = (\ref{eq:R2_transeff_i}) + (\ref{eq:R2_transeff_ii}) + (\ref{eq:R2_transeff_iii}) + (\ref{eq:R2_transeff_iv}) + (\ref{eq:R2_transeff_v}),
\]
each of which is an expectation of a product of two nuisance differences. Term (\ref{eq:R2_transeff_i}) pairs $(\pi^*_{\Gamma,0} - \hat \pi^*_\Gamma)$ with $(\fo(\tau) - \hat \f(\tau))$, since $|\Eo[\hat \f \mid Z] - \hat \E[\hat \f \mid Z]|$ is controlled by $\|\fo - \hat \f\|$; term (\ref{eq:R2_transeff_ii}) pairs $(\hat h(m) - h_0(m))$ with $(\hat g_C(k) - g_{C,0}(k))$; term (\ref{eq:R2_transeff_iii}) pairs $(\pi_{A,0} - \hat \pi_A)$, through the factor $\frac{\pi_{A,0}}{\hat \pi_A} - \frac{1-\pi_{A,0}}{1-\hat \pi_A}$ which vanishes at $\hat \pi_A = \pi_{A,0}$, with $(\g_0(t) - \hat \g(t))$, itself bounded by $\|h_0 - \hat h\|$ through $\g_0(t,W) = \hprod (1 - h_0(m))$; and terms (\ref{eq:R2_transeff_iv}) and (\ref{eq:R2_transeff_v}) pair $(\pi_{A,0} - \hat \pi_A)$ and $(\pi^*_{\Gamma,0} - \hat \pi^*_\Gamma)$ respectively with $(\fo(\tau) - \hat \f(\tau))$. Getting the supremum of all constant values and bounding them in each term, then applying the Cauchy-Schwarz inequality to each product yields

\begin{align*}
    |\R_2(\hat \eta^*, \eta_0^*; \lambda_T(\tau))| & \leq \hsumtau \sum_{k = 0}^{m -1} \Big[ \alpha_1 \|h_0(m) - \hat h(m)\| \, \|g_{C,0}(k) - \hat g_C(k)\| + \alpha_2 \|h_0(m) - \hat h(m)\| \, \|\pi_{A,0} - \hat \pi_A\| \\
    & + \alpha_3 \|\fo(\tau) - \hat \f(\tau)\| \, \|\pi_{A,0} - \hat \pi_A\| + \alpha_4 \|\fo(\tau) - \hat \f(\tau)\| \, \|\pi^*_{\Gamma,0} - \hat \pi^*_\Gamma\| \Big] \\
    & \leq \hsumtau \sum_{k=0}^{m-1} O_{\hat \P} \bigg[ \left( \|\fo(\tau) - \hat \f (\tau)\| + \| h_0(m) - \hat h(m)\| \right) \\
    & \quad \times \left( \|g_{c,0}(k) - \hat g_C(k) \| + \| \pi_{A,0} - \hat \pi_A \| + \| \pi^*_{\Gamma,0} - \hat \pi^*_{\Gamma} \| \right) \bigg]
\end{align*}
which matches equation (\ref{eq:R2_eff}).

\textbf{For $\lambda_G(\tau)$:} The remainder $\R_2(\hat \eta^*, \eta_0^*; \lambda_G(\tau))$ has the same form as $\theta_G$ with $\pi^*_{\Gamma,0}$ replacing $\pi_{\Gamma,0}$. Applying Lemma \ref{supp_lemma:seq} and grouping the propensity numerator exactly as for $\theta_G$ yields

\begin{align*}
    \R_2(\hat \eta^*, \eta_0^*; \lambda_G(\tau)) & = \hsumtau \sum_{k = 0}^{m -1} \Eo \left[ \frac{\So(m) \hat \S(t)}{\hat \S(m)} \frac{\pi_{A,0} \pi^*_{\Gamma,0} \hat \G(k) \Go(t)}{\hat \pi^*_{\Gamma} \hat \pi_{A} \hat \G(m) \Go(k)}  (\hat h(m) - h_0(m)) (\hat g_{C}(k) - g_{C,0}(k))   \right] \\
    & + \hsumtau \sum_{k = 0}^{m -1} \Eo \left[ \frac{\So(m) \hat \S(t)}{\hat \S(m)}  \frac{1}{\hat \pi_{A}}      (\hat h(m) - h_0(m)) (\pi_{A,0} - \hat \pi_A)  \right] \\
    & + \hsumtau \sum_{k = 0}^{m -1} \Eo \left[ \frac{\So(m) \hat \S(t)}{\hat \S(m)}  \frac{ \pi_{A,0}   }{\hat \pi^*_{\Gamma} \hat \pi_{A} }   (\hat h(m) - h_0(m))  ( \pi^*_{\Gamma,0} - \hat \pi^*_{\Gamma} )  \right]
\end{align*}

Getting the supremum of all constant values and bounding them in each equation, then applying the Cauchy-Schwarz inequality yields

\begin{align*}
    |\R_2(\hat \eta^*, \eta_0^*; \lambda_G(\tau))| & \leq \hsumtau \sum_{k = 0}^{m -1} \|\hat h(m) - h_0(m)\| \left( \gamma_1 \|\hat g_{C}(k) - g_{C,0}(k)\| + \gamma_2 \|\hat \pi_A - \pi_{A,0}\| + \gamma_3 \|\hat \pi^*_{\Gamma} - \pi^*_{\Gamma,0}\|  \right)
\end{align*}
Finally, since $\|h_0(m) - \hat h(m)\| \leq \|\fo(\tau) - \hat \f(\tau)\| + \|h_0(m) - \hat h(m)\|$, this is bounded by

\begin{align*}
    |\R_2(\hat \eta^*, \eta_0^*; \lambda_G(\tau))| & \leq \hsumtau \sum_{k = 0}^{m -1} O_{\hat \P} \bigg[ \left( \|\fo(\tau) - \hat \f (\tau)\| + \| h_0(m) - \hat h(m)\| \right) \\
    & \quad \times \left( \|g_{c,0}(k) - \hat g_C(k) \| + \| \pi_{A,0} - \hat \pi_A \| + \| \pi^*_{\Gamma,0} - \hat \pi^*_{\Gamma} \| \right) \bigg]
\end{align*}
which matches equation (\ref{eq:R2_eff}).


\subsection{Proof of Theorem \ref{thm:wc}}


\subsubsection{Base estimators}

For the cross-fitted estimator $\tilde{\theta}_{T,n}(t)$ we have

\begin{align*}
        \tilde{\theta}_{T,n}(t) = \sum_{b=1}^B \left( \frac{n_b}{n}  \right) \hat{\theta}_{T,n,b}(t)
\end{align*}

We have the natural one-step estimator:

\begin{equation*}
    \hat{\theta}_{T,n,b}(t) = \theta(\hat{\PP}_{-b}) + \PPn^b \left( \varphi_{T,\hat{\PP}_{-b}, t} \right)
\end{equation*}

Leading to decomposition 

\begin{align*}
    \hat{\theta}_{T,n,b}(t)- \theta_0 &= \theta(\hat{\PP}_{-b}) + \PPn^b \left( \varphi_{T,\hat{\PP}_{-b}, t} \right) - \theta(\PPo) \\
    & = \left(\PPn^b - \PPo \right) \left(  \varphi_{T,\hat{\PP}_{-b}, t} \right) + R_2 ( \hat{\PP}_{-b} , \PPo) \\
    & = \left(\PPn^b - \PPo \right) \left(  \varphi_{T,\PPo, t} \right) + \left(\PPn^b - \PPo \right) \left( \varphi_{T,\hat{\PP}_{-b}, t} - \varphi_{T,\PPo, t} \right) + R_2 ( \hat{\PP}_{-b} , \PPo) \\
    & = S^*_k + T_{1k} +T_{2k}
\end{align*}

Leading to 

\begin{align*}
    \hat{\theta}_{T,n}(t) - \theta_0 = S^* + \sum_{b=1}^B \left( \frac{n_b}{n} \right) \left( T_{1b} + T_{2b} \right) = S^* + T_1 + T_2
\end{align*}

Where:
\begin{enumerate}
    \item $S^* = \left(\PPn - \PPo \right) \left( \varphi_{T,\PPo, t} \right)$ is a sample average of fixed function, then by \textbf{Central Limit Theorem} it behaves as a normally distributed random variable with variance $\Sigma = \var \left[ \varphi_{T,\PP, t} \right]/n$ up to error $o_{\PPo} (1 / \sqrt{n})$.
    \item $T_1 = \left(\PPn - \PPo \right) \left( \varphi_{T,\hat{\PP}, t} - \varphi_{T,\PPo, t} \right)$ is the empirical process term, and is typically the smallest order since it is a sample average of a term with shrinking variance when $\varphi_{T,\hat{\PP}, t}$ converges to $\varphi_{T,\PPo, t}$.
    \item $T_2 = R_2 ( \PP , \PPo)$ is the crucial term. For non corrected plug-ins estimators this term will dominate, but for onestep estimators it will involve second order products or errors, which can be negligeable under non-parametric conditions.
\end{enumerate}

Let's again focus on a single treatment arm for simplicity, which will generalize naturally to the difference of the treatment arms. We have for term $T_2$, Theorem \ref{thm:vonmises} implies that if $\| h_0(t) - \hat h(t) \| = o_{\PP}(n^{-1/4})$ and $\| g_{C,0}(t) - \hat g_C(t) \| + \| \pi_{A,0} - \pi_A \| + \| \pi_{\Gamma,0} - \hat \pi_{\Gamma} \| = o_{\PP}(n^{-1/4})$ then $|T_2| = o_{\PP}(n^{-1/2})$. \

Recall we made the following causal assumptions. For some $c < \infty$, $P_0(1/\pi_{A,0} < c) = P_0(1/\pi_{\Gamma,0} < c) = P_0(1/\So < c) = P_0(1/\Go < c) = 1$. Additionally we assume here that $P_0(1/\hat\pi_{A,n,b} < c) = P_0(1/\hat\pi_{\Gamma,n,b} < c) = P_0(1/\hat{S}_{n,b} < c) = P_0(1/\hat{G}_{n,b} < c) = 1$. Additionally, assume that all sequences converge in probability to their true value (i.e.: $o_P(1)$), namely: $\hat\pi_{A,n,b} \overset{P}{\to} \pi_{A,0}$, $\hat\pi_{\Gamma,n,b} \overset{P}{\to} \pi_{\Gamma,0}$; $\hat{S}_{n,b}(m) \overset{P}{\to} \So(m)$; $\hat{G}_{n,b}(m) \overset{P}{\to} \Go(m)$ and $\hat{h}_{n,b}(m) \overset{P}{\to} h_0(m)$ for all $m = 1, \dots, \tau$. 

\begin{align*}
    \hat f(t) - f_0(t) & = \varphi_{T,\hat{\PP}_{-b}, t,a} - \varphi_{T,\PPo, t,a} \\
    \propto
    & = \frac{1}{\hat{P}(\Gamma = 0)}\hsum \left[  \frac{\hat{S}_{n,b}(t) \hat{\pi}_{\Gamma = 0,n,b} \left( \hat{h}_{n,b}(m) - L_m \right) }{\hat{\pi}_{A,n,b}  \hat{\pi}_{\Gamma = 1,n,b} \hat{S}_{n,b}(m-1) \hat{G}_{n,b}(m) } \right]  + \1(\Gamma = 0) \hat{S}_{n,b}(t) \\
    & - \frac{1}{\Po(\Gamma = 0)}\hsum \left[\frac{\So(t) \pi_{\Gamma=0,0} \left( h_0(m) - L_m \right) }{\pi_{A,0}  \pi_{\Gamma=1,0} \So(m-1) \Go(m) }  \right] - \1(\Gamma = 0)\So(t)
\end{align*}

Under our assumptions there is no discontinuity points in the metric space of any of the functions. Therefore, we can apply the continuous mapping theorem to function $\hat f(t) - f_0(t)$, i.e.:

\begin{align*}
    &\frac{1}{\hat{P}(\Gamma = 0)}\hsum \left[  \frac{\hat{S}_{n,b}(t) \hat{\pi}_{\Gamma = 0,n,b} \left( \hat{h}_{n,b}(m) - L_m \right) }{\hat{\pi}_{A,n,b}  \hat{\pi}_{\Gamma = 1,n,b} \hat{S}_{n,b}(m-1) \hat{G}_{n,b}(m) } \right]  + \1(\Gamma = 0) \hat{S}_{n,b}(t) \\
    & \overset{P}{\to} \frac{1}{\Po(\Gamma = 0)}\hsum \left[\frac{\So(t) \pi_{\Gamma=0,0} \left( h_0(m) - L_m \right) }{\pi_{A,0}  \pi_{\Gamma=1,0} \So(m-1) \Go(m) }  \right] - \1(\Gamma = 0)\So(t)
\end{align*}

It follows that $\hat f(t) \overset{P}{\to} f_0(t)$ and $\|\hat f(t) - f_0(t)\| = o_P(1)$. Applying Lemma \ref{supp_lemma:T1_conv} we have 

\[
T_1 = \left(\PPn^b - \PPo \right) (\hat{f}(t) - f_0(t)) = \mathcal{O}_{\PP} \left( \frac{\| \hat{f}(t) - f_0(t)\| }{\sqrt{n}} \right) = o_{\PP}(1/\sqrt{n})
\]

Finally applying Proposition \ref{supp_lemma:cv_as_1} and \ref{supp_lemma:cv_as_2} successively in combination with the Central Limit Theorem on 
\[
\tilde{\theta}_{T,n}(t) - \theta = S^* + \sum_{b=1}^B \left( \frac{n_b}{n} \right) \left( T_{1b} + T_{2b} \right)
\]

Yields the desired result: 

\[ \lim_{n \to \infty} \sqrt{n} \left( \tilde{\theta}_{T,n}(t) - \theta_0(t) \right) \underset{\text{d}}{\to} \mathcal{N} (0, \text{Var}[\varphi_{\theta_T}(O; \hat \eta) ]) \] 

Where $\Sigma = \var_{\P} \left[ \varphi_{T,\P,t} \right]$ is the non-parametric efficiency bound.

A similar argument follows for the cross-fitted estimator $\tilde{\theta}_{G,n}(t)$ we have

\begin{align*}
        \tilde{\theta}_{G,n}(t) = \sum_{b=1}^B \left( \frac{n_b}{n}  \right) \hat{\theta}_{G,n,b}(t)
\end{align*}

With natural one-step estimator:

\begin{equation*}
    \hat{\theta}_{G,n,b}(t) = \theta(\hat{\PP}_{-b}) + \PPn^b \left( \varphi_{G,\hat{\PP}_{-b}, t} \right)
\end{equation*}

Leading to the desired result:

\[ \lim_{n \to \infty} \sqrt{n} \left( \tilde{\theta}_{G,n}(t) - \theta_0(t) \right) \underset{\text{d}}{\to} \mathcal{N} (0, \text{Var}[\varphi_{\theta_G}(O; \hat \eta) ]) \] 

Where $\Sigma = \var_{\P} \left[ \varphi_{G,\P,t} \right]$ is the non-parametric efficiency bound.


\subsubsection{Structured estimators}

For the cross-fitted estimator $\tilde{\lambda}_{T,n}(t)$ we have

\begin{align*}
        \tilde{\lambda}_{T,n}(t) = \sum_{b=1}^B \left( \frac{n_b}{n}  \right) \hat{\lambda}_{T,n,b}(t)
\end{align*}

We have the natural one-step estimator:

\begin{equation*}
    \hat{\lambda}_{T,n,b}(t) = \lambda(\hat{\PP}_{-b}) + \PPn^b \left( \varphi_{T,\hat{\PP}_{-b}, t} \right)
\end{equation*}

Leading to decomposition 

\begin{align*}
    \hat{\lambda}_{T,n,b}(t)- \lambda_0 &= \lambda(\hat{\PP}_{-b}) + \PPn^b \left( \varphi_{T,\hat{\PP}_{-b}, t} \right) - \lambda(\PPo) \\
    & = \left(\PPn^b - \PPo \right) \left(  \varphi_{T,\hat{\PP}_{-b}, t} \right) + R_2 ( \hat{\PP}_{-b} , \PPo) \\
    & = \left(\PPn^b - \PPo \right) \left(  \varphi_{T,\PPo, t} \right) + \left(\PPn^b - \PPo \right) \left( \varphi_{T,\hat{\PP}_{-b}, t} - \varphi_{T,\PPo, t} \right) + R_2 ( \hat{\PP}_{-b} , \PPo) \\
    & = S^*_k + T_{1k} +T_{2k}
\end{align*}

Leading to 

\begin{align*}
    \hat{\lambda}_{T,n}(t) - \lambda_0 = S^* + \sum_{b=1}^B \left( \frac{n_b}{n} \right) \left( T_{1b} + T_{2b} \right) = S^* + T_1 + T_2
\end{align*}

Where:
\begin{enumerate}
    \item $S^* = \left(\PPn - \PPo \right) \left( \varphi_{T,\PPo, t} \right)$ is a sample average of fixed function, then by \textbf{Central Limit Theorem} it behaves as a normally distributed random variable with variance $\Sigma = \var \left[ \varphi_{T,\PP, t} \right]/n$ up to error $o_{\PPo} (1 / \sqrt{n})$.
    \item $T_1 = \left(\PPn - \PPo \right) \left( \varphi_{T,\hat{\PP}, t} - \varphi_{T,\PPo, t} \right)$ is the empirical process term, and is typically the smallest order since it is a sample average of a term with shrinking variance when $\varphi_{T,\hat{\PP}, t}$ converges to $\varphi_{T,\PPo, t}$.
    \item $T_2 = R_2 ( \PP , \PPo)$ is the crucial term. For non corrected plug-ins estimators this term will dominate, but for onestep estimators it will involve second order products or errors, which can be negligeable under non-parametric conditions.
\end{enumerate}

Let's again focus on a single treatment arm for simplicity, which will generalize naturally to the difference of the treatment arms. We have for term $T_2$, Theorem \ref{thm:vonmises} implies that if $\| \f_0(t) - \hat \f(t) \| + \| h_0(t) - \hat h(t) \| = o_{\PP}(n^{-1/4})$ and $\| g_{C,0}(t) - \hat g_C(t) \| + \| \pi_{A,0} - \pi_A \| \| \pi_{\Gamma,0} - \hat \pi_{\Gamma} \| = o_{\PP}(n^{-1/4})$ then $|T_2| = o_{\PP}(n^{-1/2})$. \

Recall we made the following causal assumptions. For some $c < \infty$, $P_0(1/\pi_{A,0} < c) = P_0(1/\pi_{\Gamma,0} < c) = P_0(1/\So < c) = P_0(1/\Go < c) = 1$. Additionally we assume here that $P_0(1/\hat\pi_{A,n,b} < c) = P_0(1/\hat\pi_{\Gamma,n,b} < c) = P_0(1/\hat{S}_{n,b} < c) = P_0(1/\hat{G}_{n,b} < c) = 1$. Additionally, assume that all sequences converge in probability to their true value (i.e.: $o_P(1)$), namely: $\hat\pi_{A,n,b} \overset{P}{\to} \pi_{A,0}$, $\hat\pi_{\Gamma,n,b} \overset{P}{\to} \pi_{\Gamma,0}$, $\hat\pi^*_{\Gamma,n,b} \overset{P}{\to} \pi^*_{\Gamma,0}$; $\hat{S}_{n,b}(m) \overset{P}{\to} \So(m)$; $\hat{G}_{n,b}(m) \overset{P}{\to} \Go(m)$, $\hat{h}_{n,b}(m) \overset{P}{\to} h_0(m)$ and $\hat{\f}_{n,b}(m) \overset{P}{\to} \f_0(m)$ for all $m = 1, \dots, \tau$. 

Recall from equation (\ref{eq:trans_eff_eif}) that for a single treatment arm $A = a$ the structured transport EIF takes the form
\begin{align*}
    \varphi_{\lambda_T}(O;\eta^*, t, a) & = \frac{1}{\p} \bigg[ \frac{\1(\Gamma = 1) (1-\pi^*_{\Gamma}) }{\pi^*_{\Gamma} }  \frac{\1(A = a)}{\pi_{A}} \D(t) \\
    & + (1-\pi^*_{\Gamma}) \left( \f(t,V) - \E \left[ \f(t,V) \mid Z \right] \right) + \1(\Gamma = 0) \left( \E \left[ \f(t,V) \mid Z \right] - \lambda_{T}(t,a) \right) \bigg]
\end{align*}
where $\D(t)$ involves $\S(t)$, $\S(m)$, $\G(m)$, $h(m)$ evaluated in the source population. The difference of the EIF evaluated at estimated versus true nuisances is therefore
\begin{align*}
    \hat f(t) - f_0(t) & = \varphi_{\lambda_T}(O;\hat{\eta}^*_{-b}, t, a) - \varphi_{\lambda_T}(O;\eta^*_0, t, a) \\
    & \propto \frac{1}{\hat \p} \bigg[ \frac{\1(\Gamma = 1) (1-\hat \pi^*_{\Gamma,n,b}) }{\hat \pi^*_{\Gamma,n,b} }  \frac{\1(A = a)}{\hat \pi_{A,n,b}} \frac{ \hat \S_{n,b}(t) \1(I_m = 1) (\hat h_{n,b}(m) - \1(L_m = 1))}{\hat \S_{n,b}(m) \hat \G_{n,b}(m)} \\
    & + (1-\hat \pi^*_{\Gamma,n,b}) \left( \hat \f_{n,b}(t,V) - \hat \E \left[ \hat \f_{n,b}(t,V) \mid Z \right] \right) + \1(\Gamma = 0) \hat \E \left[ \hat \f_{n,b}(t,V) \mid Z \right] \bigg] \\
    & - \frac{1}{\p_0} \bigg[ \frac{\1(\Gamma = 1) (1-\pi^*_{\Gamma,0}) }{\pi^*_{\Gamma,0} }  \frac{\1(A = a)}{\pi_{A,0}} \frac{ \So(t) \1(I_m = 1) (h_0(m) - \1(L_m = 1))}{\So(m) \Go(m)} \\
    & + (1-\pi^*_{\Gamma,0}) \left( \fo(t) - \Eo \left[ \fo(t) \mid Z \right] \right) + \1(\Gamma = 0) \Eo \left[ \fo(t) \mid Z \right] \bigg]
\end{align*}

Under our assumptions there are no discontinuity points in the metric space of any of the nuisance functions. Since $\hat \pi_{A,n,b} \overset{P}{\to} \pi_{A,0}$, $\hat \pi^*_{\Gamma,n,b} \overset{P}{\to} \pi^*_{\Gamma,0}$, $\hat \S_{n,b}(m) \overset{P}{\to} \So(m)$, $\hat \G_{n,b}(m) \overset{P}{\to} \Go(m)$, $\hat h_{n,b}(m) \overset{P}{\to} h_0(m)$, and $\hat \f_{n,b}(t) \overset{P}{\to} \fo(t)$ for all $m = 1,\dots,\tau$, the continuous mapping theorem applied to the expression above yields $\hat f(t) \overset{P}{\to} f_0(t)$ and hence $\|\hat f(t) - f_0(t)\| = o_P(1)$. Applying Lemma \ref{supp_lemma:T1_conv} we have

\[
T_1 = \left(\PPn^b - \PPo \right) (\hat{f}(t) - f_0(t)) = \mathcal{O}_{\PP} \left( \frac{\| \hat{f}(t) - f_0(t)\| }{\sqrt{n}} \right) = o_{\PP}(1/\sqrt{n})
\]

Finally applying Proposition \ref{supp_lemma:cv_as_1} and \ref{supp_lemma:cv_as_2} successively in combination with the Central Limit Theorem on
\[
\tilde{\lambda}_{T,n}(t) - \lambda_0 = S^* + \sum_{b=1}^B \left( \frac{n_b}{n} \right) \left( T_{1b} + T_{2b} \right)
\]

Yields the desired result:

\[ \lim_{n \to \infty} \sqrt{n} \left( \tilde{\lambda}_{T,n}(t) - \lambda_{T,\Po}(t) \right) \underset{\text{d}}{\to} \mathcal{N} (0, \text{Var}[\varphi_{\lambda_T}(O; \hat \eta^*) ]) \]

Where $\Sigma = \var_{\P} \left[ \varphi_{\lambda_T,\P,t} \right]$ is the non-parametric efficiency bound.

A similar argument follows for $\tilde{\lambda}_{G,n}(t)$. The same decomposition into $S^* + T_1 + T_2$ applies. For $T_2$, Theorem \ref{thm:vonmises} with (\ref{eq:R2_eff}) implies $|T_2| = o_\PP(n^{-1/2})$ under the same rate conditions. For $T_1$, recall from equation (\ref{eq:gen_eff_eif}) that for a single treatment arm the EIF takes the form
\begin{align*}
    \varphi_{\lambda_G}(O;\eta^*, t, a) = \frac{\1(\Gamma = 1, A=a)}{\pi^*_\Gamma \pi_A} \D(t) + \So(t \mid a, w, \gamma) - \lambda_G(t,a)
\end{align*}
The difference of the EIF at estimated versus true nuisances is
\begin{align*}
    \hat f(t) - f_0(t) & = \varphi_{\lambda_G}(O;\hat{\eta}^*_{-b}, t, a) - \varphi_{\lambda_G}(O;\eta^*_0, t, a) \\
    & \propto \frac{\1(\Gamma = 1, A=a)}{\hat \pi^*_{\Gamma,n,b} \hat \pi_{A,n,b}} \frac{ \hat \S_{n,b}(t) \1(I_m = 1) (\hat h_{n,b}(m) - \1(L_m = 1))}{\hat \S_{n,b}(m) \hat \G_{n,b}(m)} + \hat \S_{n,b}(t \mid a) \\
    & - \frac{\1(\Gamma = 1, A=a)}{\pi^*_{\Gamma,0} \pi_{A,0}} \frac{ \So(t) \1(I_m = 1) (h_0(m) - \1(L_m = 1))}{\So(m) \Go(m)} - \So(t \mid a)
\end{align*}
This has the same structure as the base generalization EIF difference but with $\pi^*_{\Gamma}(Z)$ replacing $\pi_{\Gamma}(W)$. Under our convergence assumptions on all nuisance sequences, the continuous mapping theorem yields $\|\hat f(t) - f_0(t)\| = o_P(1)$, and Lemma \ref{supp_lemma:T1_conv} gives $T_1 = o_\PP(n^{-1/2})$. Applying Propositions \ref{supp_lemma:cv_as_1} and \ref{supp_lemma:cv_as_2} yields

\[ \lim_{n \to \infty} \sqrt{n} \left( \tilde{\lambda}_{G,n}(t) - \lambda_{G,\Po}(t) \right) \underset{\text{d}}{\to} \mathcal{N} (0, \text{Var}[\varphi_{\lambda_G}(O; \hat \eta^*) ]) \]

Where $\Sigma = \var_{\P} \left[ \varphi_{\lambda_G,\P,t} \right]$ is the non-parametric efficiency bound.


\subsection{Proof of Theorem \ref{thm:drc}}

\subsubsection{Base transport estimator}

\begin{align*}
    \tilde{\theta}_{T,n}(t) & = \Eo \left[ \varphi_{T,\P,t} \right] \\
    & = \Eo \left[ \varphi_{T,\Po,t} \right] - R_2(P, \Po)(t) \\
    & = \Eo \left[ \varphi_{T,\Po,t} \right] + O_P ( R_2(P, \Po)(t) ) \\
    & = \Eo \left[ \varphi_{T,\Po,t} \right] + O_P \left[ \hsum \|h_0(m) - h(m) \| \times \left( \|g_{C,0}(m) - \hat g_C(m) \| + \|\pi_{A,0} - \hat \pi_A\| + \|\pi_{\Gamma,0} - \hat \pi_{\Gamma}\|\right) \right]
\end{align*}

If $\|h_0(m) - h(m) \| = o_P(1)$ then:

\begin{align*}
    \tilde{\theta}_{T,n}(t) & = \Eo \left[ \varphi_{T,\Po,t} \right] + O_P \left[ \hsum o_P(1) \times \left( \|g_{C,0}(m) - \hat g_C(m) \| + \|\pi_{A,0} - \hat \pi_A\| + \|\pi_{\Gamma,0} - \hat \pi_{\Gamma}\|\right) \right] \\
    & = \Eo \left[ \varphi_{T,\Po,t} \right] + o_P(1) \\
        & = \theta_{T,0}(t) + o_P(1)
\end{align*}

Similarly, if $\|g_{C,0}(m) - \hat g_C(m) \| + \|\pi_{A,0} - \hat \pi_A\| + \|\pi_{\Gamma,0} - \hat \pi_{\Gamma}\| = o_P(1)$:

\begin{align*}
    \tilde{\theta}_{T,n}(t) & = \Eo \left[ \varphi_{T,\Po,t} \right] + O_P \left[ \hsum \|h_0(m) - \hat h(m) \| \times o_P(1)  \right] \\
    & = \Eo \left[ \varphi_{T,\Po,t} \right] + o_P(1) \\
    & = \theta_{T,0 }(t) + o_P(1)
\end{align*}

Thus if either hold we have $\tilde{\theta}_{T,n}(t) = \theta_{T,0}(t) + o_P(1)$ and $\tilde \theta_{T,n}$ is a doubly-robust consistent estimator of $\theta_{T,0}$.

\subsubsection{Base generalization estimator}

The remainder $\R_2(\hat \eta, \eta_0; \theta_G(t))$ has the same product-of-errors structure as for $\theta_T$ (see equation \ref{eq:gen_R2}), so we have

\begin{align*}
    \tilde{\theta}_{G,n}(t) & = \Eo \left[ \varphi_{G,\Po,t} \right] - R_2(P, \Po)(t) \\
    & = \Eo \left[ \varphi_{G,\Po,t} \right] + O_P \left[ \hsum \|h_0(m) - \hat h(m) \| \times \left( \|g_{C,0}(m) - \hat g_C(m) \| + \|\pi_{A,0} - \hat \pi_A\| + \|\pi_{\Gamma,0} - \hat \pi_{\Gamma}\|\right) \right]
\end{align*}

If $\|h_0(m) - \hat h(m) \| = o_P(1)$ then:

\begin{align*}
    \tilde{\theta}_{G,n}(t) & = \Eo \left[ \varphi_{G,\Po,t} \right] + O_P \left[ \hsum o_P(1) \times \left( \|g_{C,0}(m) - \hat g_C(m) \| + \|\pi_{A,0} - \hat \pi_A\| + \|\pi_{\Gamma,0} - \hat \pi_{\Gamma}\|\right) \right] \\
    & = \Eo \left[ \varphi_{G,\Po,t} \right] + o_P(1) \\
    & = \theta_{G,0}(t) + o_P(1)
\end{align*}

Similarly, if $\|g_{C,0}(m) - \hat g_C(m) \| + \|\pi_{A,0} - \hat \pi_A\| + \|\pi_{\Gamma,0} - \hat \pi_{\Gamma}\| = o_P(1)$:

\begin{align*}
    \tilde{\theta}_{G,n}(t) & = \Eo \left[ \varphi_{G,\Po,t} \right] + O_P \left[ \hsum \|h_0(m) - \hat h(m) \| \times o_P(1) \right] \\
    & = \Eo \left[ \varphi_{G,\Po,t} \right] + o_P(1) \\
    & = \theta_{G,0}(t) + o_P(1)
\end{align*}

Thus if either hold we have $\tilde{\theta}_{G,n}(t) = \theta_{G,0}(t) + o_P(1)$ and $\tilde \theta_{G,n}$ is a doubly-robust consistent estimator of $\theta_{G,0}$.

\subsubsection{Structured transport estimator}

From equation (\ref{eq:R2_eff}) the remainder $\R_2(\hat \eta^*, \eta_0^*; \lambda_T(t))$ has the product-of-errors structure involving $(\fo, h_0)$ in one factor and $(g_C, \pi_A, \pi^*_\Gamma)$ in the other:

\begin{align*}
    \tilde{\lambda}_{T,n}(t) & = \Eo \left[ \varphi_{\lambda_T,\Po,t} \right] - R_2(P, \Po)(t) \\
    & = \Eo \left[ \varphi_{\lambda_T,\Po,t} \right] + O_P \bigg[ \hsum \left( \|\fo(\tau) - \hat \f(\tau)\| + \|h_0(m) - \hat h(m) \| \right) \\
    & \quad \times \left( \|g_{C,0}(m) - \hat g_C(m) \| + \|\pi_{A,0} - \hat \pi_A\| + \|\pi^*_{\Gamma,0} - \hat \pi^*_{\Gamma}\|\right) \bigg]
\end{align*}

If $\|\fo(\tau) - \hat \f(\tau)\| + \|h_0(m) - \hat h(m) \| = o_P(1)$ then:

\begin{align*}
    \tilde{\lambda}_{T,n}(t) & = \Eo \left[ \varphi_{\lambda_T,\Po,t} \right] + O_P \bigg[ \hsum o_P(1) \times \left( \|g_{C,0}(m) - \hat g_C(m) \| + \|\pi_{A,0} - \hat \pi_A\| + \|\pi^*_{\Gamma,0} - \hat \pi^*_{\Gamma}\|\right) \bigg] \\
    & = \Eo \left[ \varphi_{\lambda_T,\Po,t} \right] + o_P(1) \\
    & = \lambda_{T,0}(t) + o_P(1)
\end{align*}

Similarly, if $\|g_{C,0}(m) - \hat g_C(m) \| + \|\pi_{A,0} - \hat \pi_A\| + \|\pi^*_{\Gamma,0} - \hat \pi^*_{\Gamma}\| = o_P(1)$:

\begin{align*}
    \tilde{\lambda}_{T,n}(t) & = \Eo \left[ \varphi_{\lambda_T,\Po,t} \right] + O_P \bigg[ \hsum \left( \|\fo(\tau) - \hat \f(\tau)\| + \|h_0(m) - \hat h(m) \| \right) \times o_P(1) \bigg] \\
    & = \Eo \left[ \varphi_{\lambda_T,\Po,t} \right] + o_P(1) \\
    & = \lambda_{T,0}(t) + o_P(1)
\end{align*}

Thus if either hold we have $\tilde{\lambda}_{T,n}(t) = \lambda_{T,0}(t) + o_P(1)$ and $\tilde \lambda_{T,n}$ is a doubly-robust consistent estimator of $\lambda_{T,0}$.

\subsubsection{Structured generalization estimator}

The remainder $\R_2(\hat \eta^*, \eta_0^*; \lambda_G(t))$ has the same product-of-errors structure as for $\lambda_T$ (see equation \ref{eq:R2_eff}):

\begin{align*}
    \tilde{\lambda}_{G,n}(t) & = \Eo \left[ \varphi_{\lambda_G,\Po,t} \right] - R_2(P, \Po)(t) \\
    & = \Eo \left[ \varphi_{\lambda_G,\Po,t} \right] + O_P \bigg[ \hsum \left( \|\fo(\tau) - \hat \f(\tau)\| + \|h_0(m) - \hat h(m) \| \right) \\
    & \quad \times \left( \|g_{C,0}(m) - \hat g_C(m) \| + \|\pi_{A,0} - \hat \pi_A\| + \|\pi^*_{\Gamma,0} - \hat \pi^*_{\Gamma}\|\right) \bigg]
\end{align*}

If $\|\fo(\tau) - \hat \f(\tau)\| + \|h_0(m) - \hat h(m) \| = o_P(1)$ then:

\begin{align*}
    \tilde{\lambda}_{G,n}(t) & = \Eo \left[ \varphi_{\lambda_G,\Po,t} \right] + O_P \bigg[ \hsum o_P(1) \times \left( \|g_{C,0}(m) - \hat g_C(m) \| + \|\pi_{A,0} - \hat \pi_A\| + \|\pi^*_{\Gamma,0} - \hat \pi^*_{\Gamma}\|\right) \bigg] \\
    & = \Eo \left[ \varphi_{\lambda_G,\Po,t} \right] + o_P(1) \\
    & = \lambda_{G,0}(t) + o_P(1)
\end{align*}

Similarly, if $\|g_{C,0}(m) - \hat g_C(m) \| + \|\pi_{A,0} - \hat \pi_A\| + \|\pi^*_{\Gamma,0} - \hat \pi^*_{\Gamma}\| = o_P(1)$:

\begin{align*}
    \tilde{\lambda}_{G,n}(t) & = \Eo \left[ \varphi_{\lambda_G,\Po,t} \right] + O_P \bigg[ \hsum \left( \|\fo(\tau) - \hat \f(\tau)\| + \|h_0(m) - \hat h(m) \| \right) \times o_P(1) \bigg] \\
    & = \Eo \left[ \varphi_{\lambda_G,\Po,t} \right] + o_P(1) \\
    & = \lambda_{G,0}(t) + o_P(1)
\end{align*}

Thus if either hold we have $\tilde{\lambda}_{G,n}(t) = \lambda_{G,0}(t) + o_P(1)$ and $\tilde \lambda_{G,n}$ is a doubly-robust consistent estimator of $\lambda_{G,0}$.
    
\section{Web Appendix C: Numerical Study
(Section \ref{sec:num})} \label{sec:supp_num}

\subsection{Data-generating mechanism \ref{sec:num_DMG}}\label{sec:supp_num_DMG}

Our numerical study requires a data-generating mechanism that is compatible with the structural assumptions underpinning the structured subset estimands. Designing a simulation that simultaneously respects these conditions is nontrivial. In discrete time, the survival function is linked to the hazard through
\[
S(t \mid A,W,\Gamma) \;=\; \prod_{m=1}^t \bigl\{1 - h(m \mid A,W,\Gamma)\bigr\},
\]
so any specification of the hazard $h$ implicitly determines $S$, and vice versa. If one attempts to directly specify $h(m \mid A,W,\Gamma)$ to satisfy Assumptions~\ref{ass:param}--\ref{ass:tem}, it is easy to violate monotonicity of $S$ or to inadvertently reintroduce dependence of $f$ on non-effect-modifying components of $W$. Instead, we adopt a constructive procedure that enforces the assumptions by projection.

\paragraph{Step 1: Construct temporary hazards and survival.}
We begin by specifying \emph{temporary} treatment-specific hazards
\[
h_a^{\mathrm{temp}}(m \mid W), \qquad a \in \{0,1\},\; m = 1,\ldots,K,
\]
which are allowed to depend on the full covariate vector $W$ and need not satisfy Assumptions~\ref{ass:param}--\ref{ass:tem}. From these we define the corresponding temporary survival functions
\[
S_a^{\mathrm{temp}}(t \mid W)
    \;=\; \prod_{m=1}^t \bigl\{1 - h_a^{\mathrm{temp}}(m \mid W)\bigr\},
    \qquad t = 1,\ldots,K,\; a \in \{0,1\}.
\]
The associated preliminary treatment contrast on the survival scale is then
\[
f_{\mathrm{temp}}(t,W)
    \;=\; S_1^{\mathrm{temp}}(t \mid W) - S_0^{\mathrm{temp}}(t \mid W),
    \qquad t = 1,\ldots,K.
\]
By construction, $f_{\mathrm{temp}}(t,W)$ may depend on all components of $W$ and therefore does not yet enforce Assumption~\ref{ass:tem}.

\paragraph{Step 2: Projecting the contrast onto the effect-modifier space.}
To impose that the treatment contrast depends only on $V = V(W)$, we project the preliminary contrast $f_{\mathrm{temp}}(t,W)$ onto the $\sigma(V)$-field. Formally, for each $t$ we define
\[
f(t,V)
    \;=\; \mathbb{E}\bigl[ f_{\mathrm{temp}}(t,W) \,\bigm|\, V \bigr],
\]
so that $f(t,V)$ is the conditional expectation of $f_{\mathrm{temp}}(t,W)$ given the lower-dimensional effect modifiers $V$. In practice, this conditional expectation is approximated via a flexible regression of $f_{\mathrm{temp}}(t,W)$ on $V$ (e.g.\ using a Super Learner), but conceptually it is a projection of $f_{\mathrm{temp}}$ onto the space of functions of $V$ only. This step enforces Assumption~\ref{ass:tem} by construction, since $f$ no longer depends on the components of $W$ outside of $V$.

\paragraph{Step 3: Reconstructing survival functions consistent with the parameterization.}
We next rebuild the treatment-specific survival functions so that the additive parameterization in Definition~\ref{ass:param} holds. We take the temporary control survival $S_0^{\mathrm{temp}}(t \mid W)$ as a baseline component and set
\[
g(t,W) \;=\; S_0^{\mathrm{temp}}(t \mid W),
\]
so that the control arm corresponds to $A=0$. We then define the treated survival via
\[
S_0(t \mid W)
    \;=\; \mathrm{monotone}\bigl( S_0^{\mathrm{temp}}(t \mid W) \bigr), 
    \qquad
S_1(t \mid W)
    \;=\; \mathrm{monotone}\bigl( S_0^{\mathrm{temp}}(t \mid W) + f(t,V) \bigr),
\]
where $\mathrm{monotone}(\cdot)$ denotes element wise clipping to the interval $[0,1]$ followed by a cumulative minimum in $t$ to enforce non-increasing survival. The final arm-specific hazards are obtained from these survival curves via
\[
h_a(m \mid W)
    \;=\; 1 - \frac{S_a(m \mid W)}{S_a(m-1 \mid W)},
    \qquad S_a(0 \mid W) := 1,\;\; a \in \{0,1\}.
\]
With this construction, the observed-arm survival can be written as
\[
S(t \mid A,W,\Gamma)
    \;=\;
    \begin{cases}
        S_0(t \mid W), & A=0,\\[3pt]
        S_1(t \mid W), & A=1,
    \end{cases}
\]
which implements the desired decomposition
\[
S(t \mid A,W,\Gamma)
    \;=\; A\, f(t,V) + g(t,W).
\]
Thus Definition~\ref{ass:param} is satisfied with $f(t,V)$ obtained from the projection step and $g(t,W) = S_0(t \mid W)$.

\paragraph{Step 4: Encoding partial study heterogeneity.}
Finally, Assumption~\ref{ass:hem} concerns the joint distribution of $(\Gamma, V, Z)$ rather than the outcome model itself. In the simulation, we first generate baseline covariates $W$, then construct $V = V(W)$ and $Z \subseteq V$, and specify the population assignment mechanism $P(\Gamma = 1 \mid Z)$ so that $\Gamma$ depends on $W$ only through $Z$. This ensures $\Gamma \perp V \mid Z$ by design. The outcome model constructed in Steps 1--3 is then applied identically across populations, so that differences in marginal survival between $\Gamma=0$ and $\Gamma=1$ arise solely from differences in the distributions of the effect modifiers captured by $Z$.

Taken together, this four-step procedure yields a coherent data-generating mechanism in which (i) the survival function admits the additive parameterization $S(t \mid A,W,\Gamma) = A f(t,V) + g(t,W)$, (ii) the treatment effect $f$ depends only on the lower-dimensional effect modifiers $V$, and (iii) partial study heterogeneity is encoded through $Z$ in the sense of Assumption~\ref{ass:hem}.

\section{Web Appendix D: Additional results}
\label{sec:supp_addresults}

\begin{lemma}\label{supp_lemma:seq}
    (From \citep{diaz2018targeted}). For two sequences $a_1, \dots, a_m$ and $b_1, \dots, b_m$ we have
    \[\prod_{t=1}^m (1-a_t) - \prod_{t=1}^m (1-b_t) = \sum_{t=1}^m \left( \prod_{k=1}^{t-1}(1-a_k)(b_t-a_t) \prod_{k=t+1}^m (1-b_k)   \right) \]
\end{lemma}


\begin{lemma}\label{supp_lemma:T1_conv}
    (From \citep{kennedy2023semiparametricdoublyrobusttargeted}). Let $f(\hat{z})$ be a function estimated from a sample $Z^N = (Z_{n+1},\dots, Z_N)$, and let $\PPn$ denote the empirical measure over $(Z_{1},\dots, Z_n)$, which is independent of $Z^N$. Then 
    \[(\PPn - \PP) (\hat{f} - f) = O_{\PP} \left( \frac{\| \hat{f} - f \|}{\sqrt{n}}  \right) \]
\end{lemma}


\begin{prop}\label{supp_lemma:cv_as_1}
    (From \citep{kennedy2023semiparametricdoublyrobusttargeted}). Let $\hat{\psi} = \hat{\psi}_k (\hat{\PP}_{-k} + \PP_n^k \left\{ \varphi(Z; \hat{\PP}_{-k} \right\}$ the usual one-step estimator in the $k$th fold. Assume $K \leq C < \infty$ is finite, and that $\|\varphi(z; \hat{\PP}_{-k} - \varphi(z;\PP)\| = o_{\PP}(1)$ for each $k$. Then \[\hat{\psi} - \psi = (\PPn - \PP) \{ \varphi (Z; \PP) \} + T_2 + o_{\PP} (1/\sqrt{n}) \]
    for $T_2 = \sum_{k=1}^K \left( \frac{N_k}{n} \right) R_2(\hat{\PP}_{-k} ; \PP) $ and $R_2(\hat{P}, P) = \psi(\hat{P}) - \psi(P) + \int \varphi(z; \hat{P}) dP(z)$.
\end{prop}


\begin{prop}\label{supp_lemma:cv_as_2}
    (From \citep{kennedy2023semiparametricdoublyrobusttargeted}). Let $\hat{\psi} = \hat{\psi}_k (\hat{\PP}_{-k} + \PP_n^k \left\{ \varphi(Z; \hat{\PP}_{-k} \right\}$ the usual one-step estimator in the $k$th fold. Assume $K \leq C < \infty$ is finite, and that $\|\varphi(z; \hat{\PP}_{-k} - \varphi(z;\PP)\| = o_{\PP}(1)$ for each $k$, and that $T_2 = \sum_{k=1}^K \left( \frac{N_k}{n} \right) R_2(\hat{\PP}_{-k} ; \PP) = o_{\PP} (1/\sqrt{n})$, where $R_2(\hat{P}, P) = \psi(\hat{P}) - \psi(P) + \int \varphi(z; \hat{P}) dP(z)$. Then 
    \[\hat{\psi} - \psi = (\PPn - \PP) \{ \varphi (Z; \PP) \} + o_{\PP} (1/\sqrt{n}) \]
    and so $\hat{\psi}$ is root-n consistent, asymptotically normal, and minimax optimal in the local asymptotic if $\varphi$ is the efficient influence function. 
\end{prop}

\label{lastpage}
\end{document}